\documentclass[fleqn,usenatbib]{mnras}

\usepackage{newtxtext,newtxmath}
\usepackage[T1]{fontenc}
\usepackage{ae,aecompl}

\usepackage{graphicx}	        % Including figure files
\usepackage{amsmath}	        % Advanced maths commands
\usepackage{amssymb}	        % Extra maths symbols
\usepackage{gensymb}            % Generic symbols
\usepackage{multirow}           % Multiple rows
\usepackage{ulem}               % Strike through text
\usepackage[dvipsnames]{xcolor} % Colors
\newcommand{\tabletext}{\color{black}} % for table footnote
\title[MW mass with \textit{Gaia}]{The mass of our Galaxy from satellite proper motions in the \textit{Gaia} era}
\author[Tobias K. Fritz et al.]
    {T. K. Fritz$^{1,2}$\thanks{tfritz@iac.es}, A. Di Cintio$^{1,2}$\thanks
    {Marie-Sk\l{}odowska-Curie Fellow}, G. Battaglia$^{1,2}$, C. Brook$^{1,2}$, 
     S. Taibi$^{1,2}$\\
$^{1}$Instituto de Astrof\'{i}sica de Canarias, Calle Via L\'{a}ctea s/n, E-38206 La Laguna, Tenerife, Spain\\
$^{2}$Universidad de La Laguna. Avda. Astrof\'{i}sico Fco. S\'{a}nchez, La Laguna, Tenerife, Spain\\
}
\date{Accepted XXX. Received YYY; in original form ZZZ}

\pubyear{2020}
\begin{document}
\label{firstpage}
\pagerange{\pageref{firstpage}--\pageref{lastpage}}
\maketitle

\begin{abstract} 
 We use \textit{Gaia} DR2 systemic proper motions of 45 satellite galaxies to constrain the mass of the Milky Way using the scale free mass estimator of Watkins et al. (2010). We first determine the anisotropy parameter $\beta$, and the tracer satellites' radial density index $\gamma$ to be  $\beta$=$-0.67^{+0.45}_{-0.62}$  and
 $\gamma=2.11\pm0.23$. When we exclude possible former satellites of the Large Magellanic Cloud, the anisotropy changes to $\beta$=$-0.21^{+0.37}_{-0.51}$.
We find that the index of the Milky Way's  gravitational potential $\alpha$, which is dependent on the mass itself, is the parameter with the largest impact on the mass determination. 
Via comparison with cosmological simulations of Milky Way-like galaxies, we carried out a detailed analysis of the estimation of the  observational uncertainties and their impact on the mass estimator. We found that the mass estimator is biased when applied naively to the satellites of simulated Milky Way halos. Correcting for this bias, we  obtain for our Galaxy a  mass  of $0.58^{+0.15}_{-0.14}\times10^{12}$M$_\odot$ within 64 kpc, as computed from the inner half of our observational sample, and $1.43^{+0.35}_{-0.32}\times10^{12}$M$_\odot$ within 273 kpc, from the full sample; this latter value extrapolates to  
a virial mass of $M_\mathrm{vir\,\Delta=97}$$=$$1.51^{+0.45}_{-0.40} \times 10^{12}M_{\odot}$ corresponding to a virial radius of  R$_\mathrm{vir}$$=$$308\pm29$ kpc.
This value of the Milky Way mass lies in-between other mass estimates reported in the literature, from  various different methods.
\end{abstract}

%%%%%%%%%%%%%%%%%%%%%%%%%%%%%%%%%%%%%%%%%%%%%%%%%%%
%%%%%%%%%%%%%%%%% BODY OF PAPER %%%%%%%%%%%%%%%%%%%
%%%%%%%%%%%%%%%%%%%%%%%%%%%%%%%%%%%%%%%%%%%%%%%%%%%
\section{Introduction} 
\label{sec:intro}
%%%%%%%%%%%%%%%%%%%%%%%%%%%%%%%%%%%%%%%%%%%%%%%%%%%
In a $\Lambda$CDM universe,  galaxies are embedded in a dark matter 
halo \citep{white78}, which is the most important component in terms 
of mass. The radial density of dark matter halos can be approximated 
by a NFW profile \citep{Navarro_97},
characterised by two parameters, e.g. its virial mass and 
concentration.
Observationally, the virial mass of dark matter halos surrounding 
galaxies is difficult to obtain,  since it is best measured at large 
distances from the centre of the galaxy itself, at radii where the 
number of dynamical tracers is low. 

Having a good estimate of the MW virial mass would allow for more direct comparisons with cosmological simulations, specifically for those properties that are dependent on mass. 
For example, cosmological DM-only simulations have a number of satellites that have too much mass in the inner regions to be consistent with the observed internal kinematic properties of MW satellite galaxies . This issue, dubbed the `Too big to fail' 
 problem \citep{Boylan-Kolchin_11b}, would be solved trivially if the mass of the MW was relatively low \citep{Wang_12a,Vera-Ciro_13}. 
 
 Many methods employed so far for measuring the virial mass of the MW require extrapolation, because  the dynamical tracers either do not extend out to the virial radius (e.g. globular clusters, see \citet{Harris_96}), or  their apparent magnitude at that distance makes observations with current facilities challenging (e.g. blue horizontal branch stars). 
 The  most easily accessible tracers that cover the largest radial range are dwarf galaxies. However, past attempts to use dwarf galaxies have resulted in a range of derived halo masses, which are particularly sensitive to which galaxies are included \citep[especially Leo~I][]{Kulessa_92,Watkins_10,Boylan-Kolchin_13}: these measurements provide a total MW mass that ranges from 0.5 to 3$\times10^{12}$ M$_\odot$ \citep[see][for reviews on the MW mass]{Wang_15a,Bland_16}. A limitation of these previous studies was that  systemic proper motions were only available for a handful of MW satellite galaxies, with the consequence that the anisotropy parameter of the tracer population, $\beta$, could not be well constrained. The degeneracy between mass and anisotropy \citep[see e.g.][]{Binney_08} then leads to significant uncertainties in the mass estimation.

Recently, \textit{Gaia} DR2 \citep{Brown_18} provided proper motions of more than 1 billion stars in our Galaxy and sub-systems within, such as globular clusters and satellites. These data have been used for quantitative determinations of the MW mass based on globular clusters (see e.g. \citealt{ Watkins_18}, \citealt{Vasiliev_19}) and  fast halo stars (see e.g. \citealt{Monari_18}, \citealt{Deason_19}).  Gaia DR2 systemic proper motions for more than 40 of the MW satellites have now been obtained (see e.g. \citealt{Helmi_18}, \citealt{Fritz_18a}, \citealt{Simon_18}, \citealt{Pace_19}), 
the first such measurements for the faintest dwarfs. \citet{Callingham_19} considered the sample of classical dwarfs for a determination of the Milky Way mass, but for these systems previous HST based proper motions were already of similar precision \citep{Patel_18}. 

In this work we use the \textit{Gaia} DR2 proper motions of the full set of satellite dwarf galaxies to determine the virial mass of the Milky Way. For this purpose we adopt the \citet{Watkins_10} (Wa10) mass estimator, tailored to systems with full phase-space information. This mass estimator has the advantage of allowing arbitrary constant values for the anisotropy parameter, an arbitrary power-law mass profile for the underlying halo and an arbitrary power-law for the tracer distribution, without hidden assumptions on these quantities. Its applicability to the Milky Way has been previously tested using constrained cosmological simulations of the  Local Group of galaxies \citep{Dicintio_12}.

Other methods, like Jeans and Schwarzschild \citep{Schwarzschild_79} modelling, allow better constraints on the mass profile, but they contain a significant number of free parameters, making them difficult to apply to a still relatively small sample. Further,  the reliance on  simulations is minimal in our methodology, less than when simulations are directly used, as in \citet{Patel_18}  or \citet{Callingham_19}.

The paper is structured as follows: in Section~\ref{section:data} we describe the data and sample used; in Section~\ref{sec:abg} we derive the parameters used in the Wa10 mass estimator; in Section~\ref{sec:simulations} we compare with simulations to assess the biases of the estimator. An additional complication is given by the presence of a close-by and relatively massive neighbour, the Large Magellanic Cloud (LMC), which exerts a strong gravitational pull \citep{Gomez_15, Garavito_19}. In Section~\ref{sec:lmc} we present our assumptions regarding the reflex motion of the MW due to the Large Magellanic Cloud (LMC) and explore the effect of galaxies in the sample that are possible former satellites of the LMC. In  Section~\ref{sec:MW-mass} we derive the mass (profile) of the Milky Way and compare it with other determinations. Finally, in Section~\ref{sec:conclusions}, we conclude and summarise our results. 
%%%%%%%%%%%%%%%%%%%%%%%%%%%%%%%%%%%%
\section{Data} 
\label{section:data}
%%%%%%%%%%%%%%%%%%%%%%%%%%%%%%%%%%%%

Our sample consists of all satellite galaxies within the virial radius of the MW for which  systemic proper motions and line of sight velocities are available, which minimises the dependence on the anisotropy parameter, $\beta$. We include satellites galaxies from \citet{Fritz_18b}, \citet{Carlin_18}, \citet{Helmi_18}, \citet{Torrealba_18b}, and \citet{Longeard_19}. Details of the 
 sources of proper motions, line-of-sight velocities and distance moduli are listed in Appendix~\ref{sec:ap}. We exclude Laevens~1/Crater~I, as  it  is likely a globular cluster (see \citealt{Laevens_14}, \citealt{Kirby_15}, \citealt{Voggel_16}, \citealt{Weisz_16}). We also exclude Phoenix~I and Eridanus~II because their large distances imply that they are likely outside the virial radius of the MW. 

The proper motions are  primarily based on \citet{Fritz_18a}, which relied on samples of individual stars for which spectroscopic measurements were available. This reduces the likelihood that a proper motion measurement is affected by membership uncertainties.  We assume systematic uncertainties
of 0.035 mas/yr \citep{Helmi_18} in the
 proper motion for Bootes~III, Antlia~II and
Sagr~II since these were not included in the original sources.

\begin{figure}
\includegraphics[width=0.70\columnwidth,angle=-90]{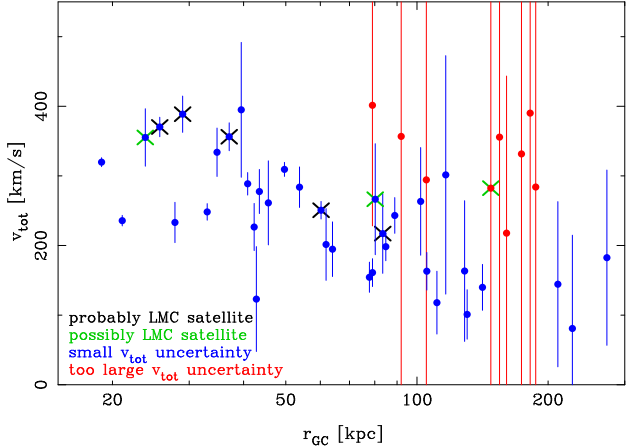} 
    \caption{Total velocity  (i.e. amplitude of the velocity vector) versus Galactocentric distance for the sample of 45 galaxies. 
    In red we indicate those galaxies that we exclude from the final sample used for the mass determination, due to the very large uncertainty in their total velocity (see Section~\ref{sec:obs-errrors} and Fig.~\ref{fig:mas_sim_err} for details). The crosses show the galaxies that are likely to be former LMC satellites.}
    \label{fig:sample}
\end{figure}

In total, our sample contains 45 satellite galaxies, as shown in Fig.~\ref{fig:sample}. In the following, we also consider  sub-samples of these 45 galaxies based on the uncertainty in the amplitude of the velocity vector, and on the possible origin as a former satellites of the LMC. These sub-samples will be defined in detail in the appropriate sections but Fig.~\ref{fig:sample} can be used for an overview.\footnote{In principle, a few objects could have been satellites of the Small Magellanic Cloud (SMC), but since the SMC was itself very likely a satellite of the LMC, we do not distinguish between the two cases.} 

To  test for the radial mass distribution and consistency of our results, we also divide the sample into two bins: an inner one, which includes satellites within 64 kpc (with the outermost object being Bootes~I), and an outer one with satellites found beyond 77 kpc (with the innermost one being Ursa Minor). With such division we achieve roughly equal sample sizes and also ensure that the inner and outermost galaxies are independent of LMC satellite selection.

%%%%%%%%%%%%%%%%%%%%%%%%%%%%%%%%%%%%
\section{Determination of $ \alpha$, $\beta$ and $\gamma$ parameters} 
\label{sec:abg}
%%%%%%%%%%%%%%%%%%%%%%%%%%%%%%%%%%%%

When galactocentric distances and total velocities are available for a sample of tracers, the Wa10 mass estimator gives the total mass of a galaxy within the radius of its outermost tracer with the equation:

\begin{equation}
M_\mathrm{<=out}=\frac{1}{G} \frac{\alpha+\gamma-2\beta}{3-2\beta}r_\mathrm{GC\,out}^{1-\alpha}  \langle v_\mathrm{tot}^2r_\mathrm{GC}^\alpha \rangle 
\label{eq:wa10_orig}
\end{equation}

\noindent where $r_\mathrm{GC}$ is the Galactocentric radius of the tracer ($r_\mathrm{GC\,out}$ being the radius of the outer most one), $v_\mathrm{tot}$ is the total velocity of the tracers and  $\alpha$, $\beta$ and $\gamma$ are, respectively, the parameters describing the radial mass distribution of the host galaxy, the anisotropy parameter (assumed to be independent of radius) and the radial distribution of the tracer population (here the MW satellite galaxies). The estimator assumes that the tracer population has a number density that follows a power-law and moves in a scale-free potential.

We derive the parameters $\alpha$, $\beta$ and $\gamma$ for the MW in the following sections.  In addition, we also calculate their values in simulations for comparison and in order to correct for biases in the estimates of the MW mass.

\subsection{Determination of the mass index $\alpha$}
\label{section:alpha}

The mass profile can be written as: 
$$ M(r) \propto r_\mathrm{GC}^{1-\alpha}$$

\noindent where $\alpha$ can be obtained by  fitting the full observational data set of tracers at every radii, or by simply interpolating between the inner most and the outer most point of the mass distribution. 
Given the assumption of a power-law, we decided to use this last method\footnote{We nevertheless verified, using the simulations, that the two methods provide the same value of $\alpha$ within 0.02.}: we therefore need to calculate the mass enclosed within the innermost and outermost radius of the objects in our sample. 

\begin{figure}
    \includegraphics[ width=0.70\columnwidth,angle=-90]{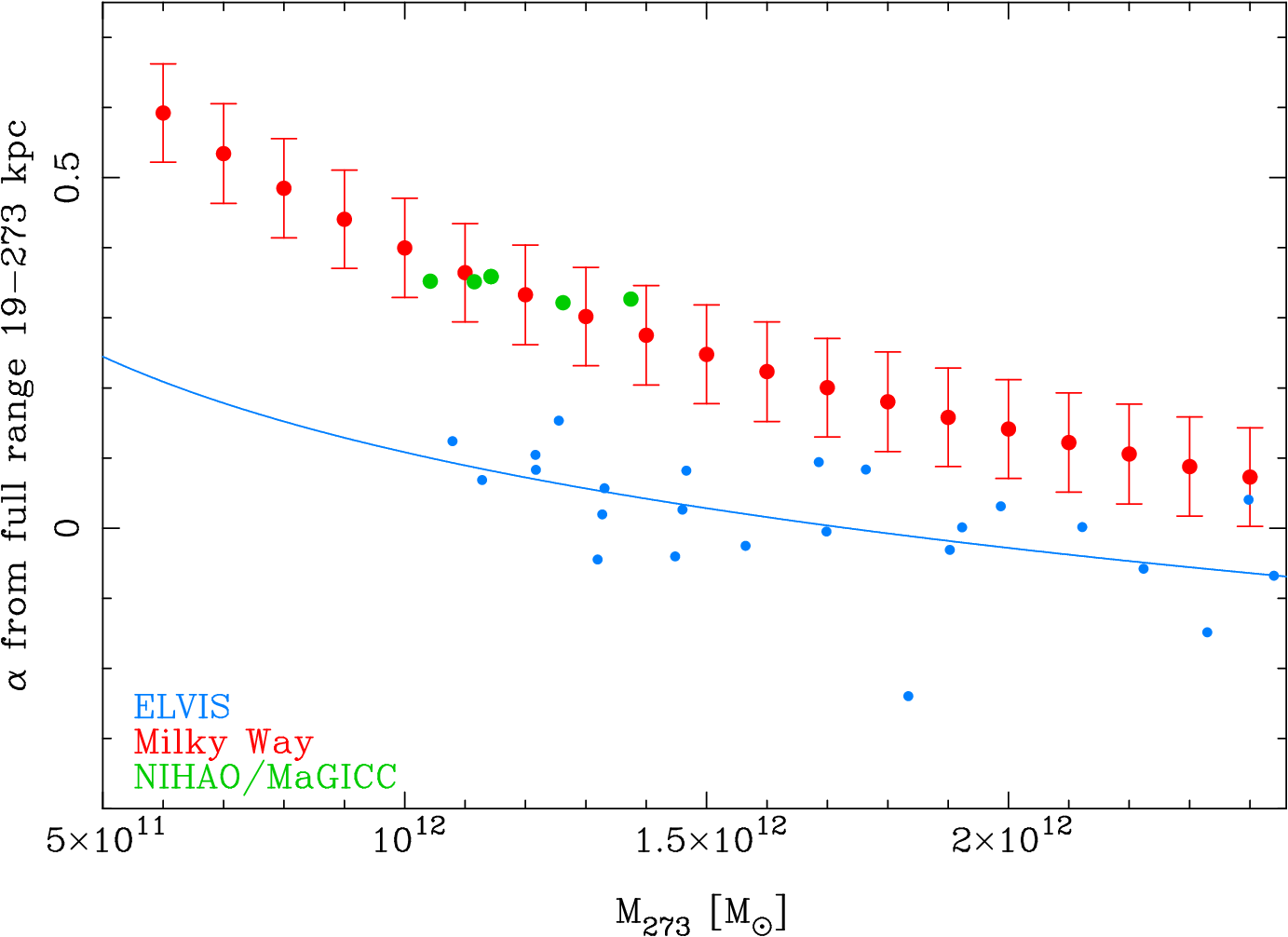}
    \includegraphics[width=0.70\columnwidth,angle=-90]{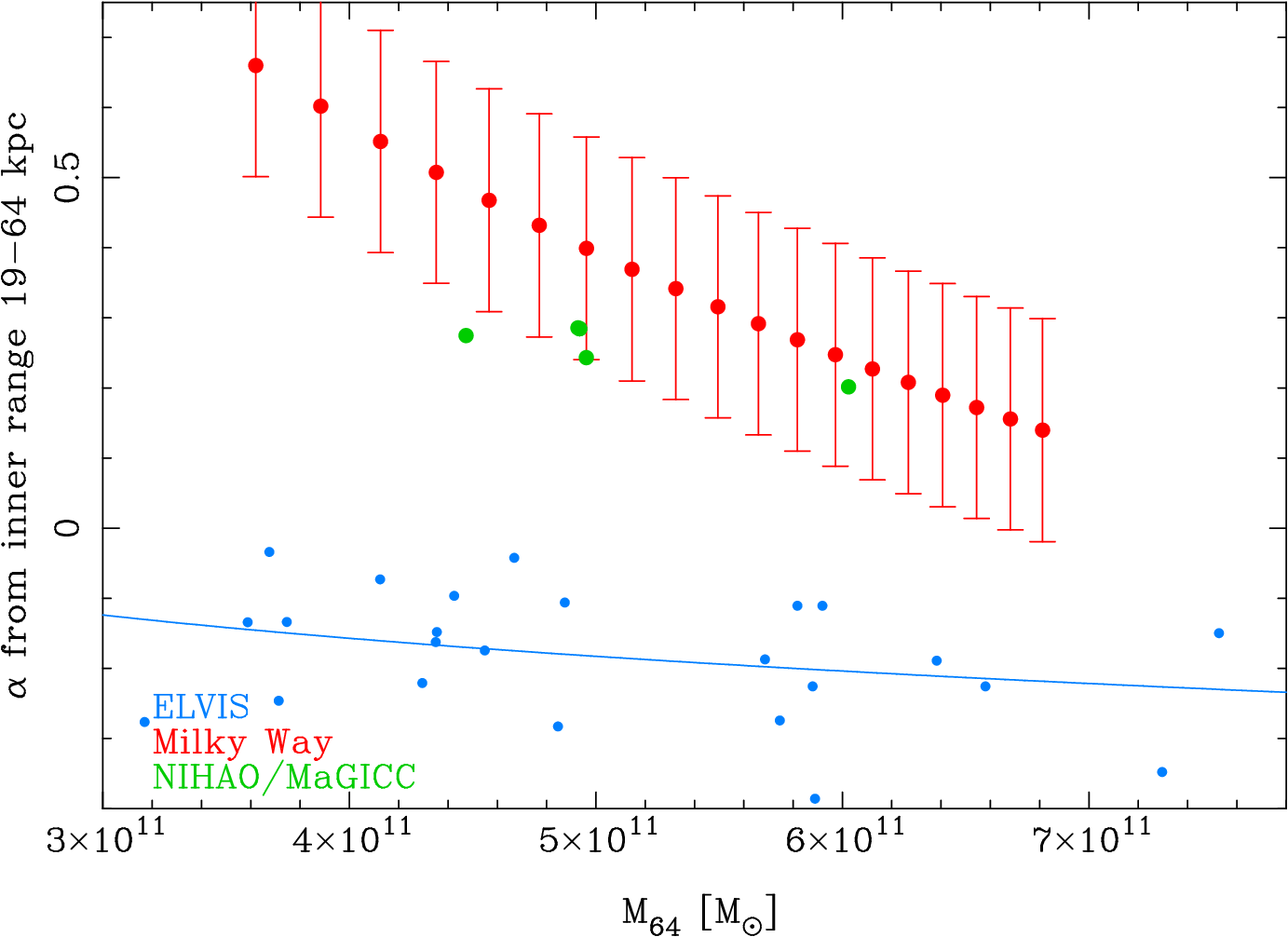}
\includegraphics[width=0.70\columnwidth,angle=-90]{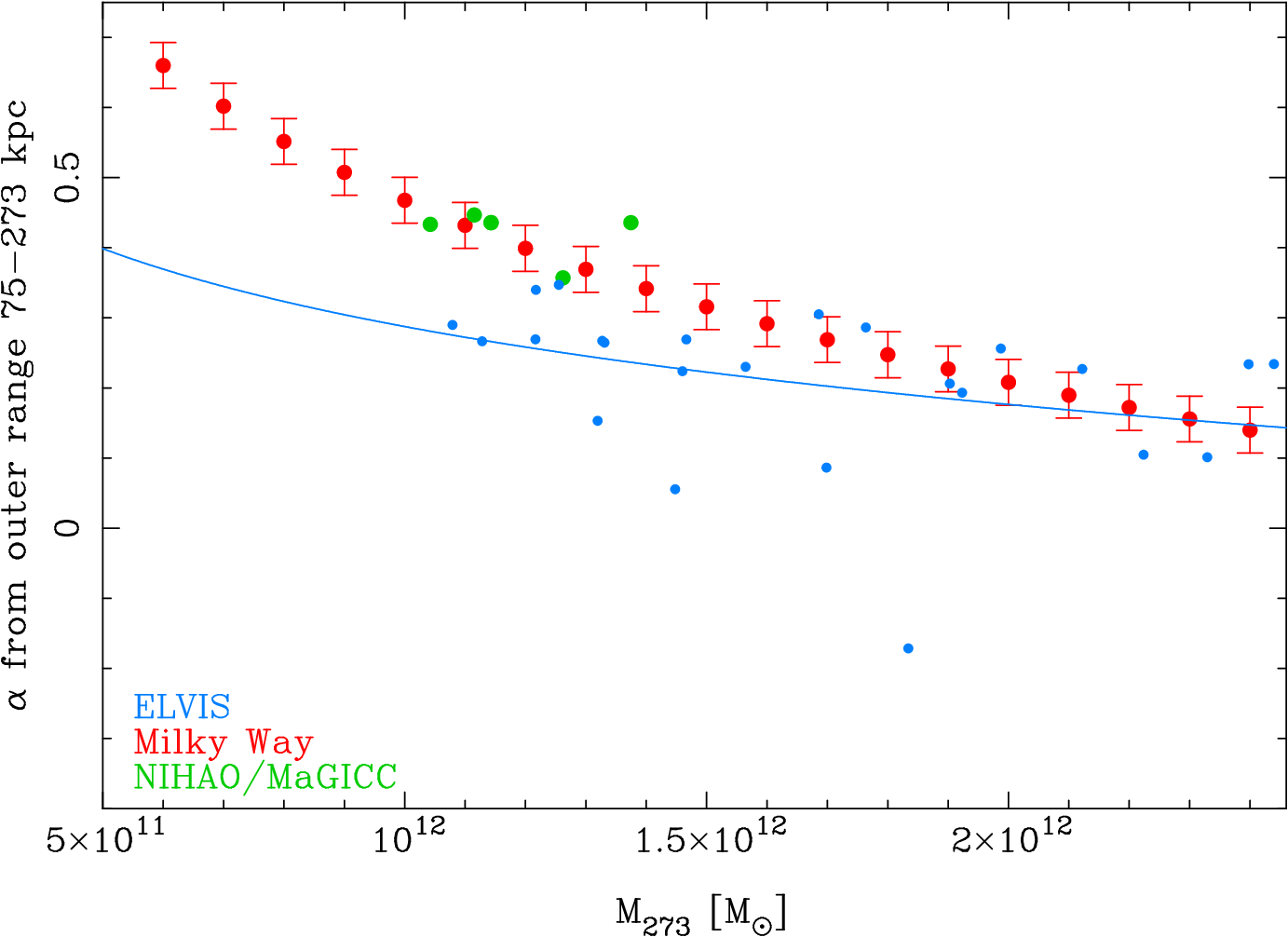}
    \caption{$\alpha$ as function of assumed mass at different outer radii (red circles with error-bars). We also show $\alpha$ obtained in hydrodynamical NIHAO/MaGICC simulations, in green, using only those simulated galaxies that have a v$_\mathrm{circ}$ at 19 kpc within 215$\pm$20 km/s,  and in dark matter only (ELVIS) simulations, as blue circles, where the solid line is a fit to the points. From top to bottom, the panels show results for the full radial range (19-273 kpc), and the inner (19-64 kpc) and outer (77-273 kpc) sample.}
    \label{fig:m_alpha}
\end{figure}

Estimates of the MW mass within the inner radius of our tracer sample ($\approx20$ kpc) agree relatively well in the literature. For example, \citet{Kuepper_15} obtained  V$_\mathrm{circ}=217^{+22}_{-19}$ km/s at $\approx19$ kpc using the Pal~5 stream; \citet{Bovy_16} obtained V$_\mathrm{circ}=198\pm9$ km/s at 20 kpc using the Pal~5 and GD1 streams;  \citet{Watkins_18} obtained V$_\mathrm{circ}=213^{+20}_{-17}$ km/s at 21.1 kpc using globular cluster dynamics. We average these measurements and obtain V$_\mathrm{circ}=215\pm20$ km/s, and we then convert such circular velocity to  a mass at 19 kpc. 
We then calculate $\alpha$ as a function of different assumed values of the mass within the outermost point ($r_{\rm GC}= $273 kpc), keeping the value of the mass within 19 kpc fixed. The results are shown in the top panel of Fig.~\ref{fig:m_alpha} as red points with error-bars: the variation of $\alpha$ with outer mass is  well fit by the quadratic function\footnote{As elsewhere in this paper $\log$ is in base 10.}:

\begin{equation}
\alpha=11.69-1.016\times\log{M_\mathrm{273}}+0.0063\times(\log{M_\mathrm{273}})^2
\label{eq:mass_alpha_full}
\end{equation}

We estimate an uncertainty in $\alpha$ of 0.07, taking into account the uncertainty of the mass at 19 kpc. Eq.~\ref{eq:mass_alpha_full} will be later used in Section~\ref{sec:MW-mass} to iteratively calculate  $\alpha$ and its associated mass value.
Following a similar procedure we also calculate $\alpha$ as a function of mass for two other radial ranges,  respectively the inner (between 19 and 64 kpc) and outer (between 77 and 273 kpc) range. The results are shown respectively in the central and bottom panels of Fig.~\ref{fig:m_alpha}, as red points with error-bars.
Once again, a quadratic relation is used to compute the variation of $\alpha$ with mass:
for the inner range, it reads as 
$\alpha=20.43-1.554\times\log{M_\mathrm{64}}-0.0146\times(\log{M_\mathrm{64}})^2$ 
with an uncertainty on $\alpha$ of 0.159, while for the outer range we get $\alpha=11.66-1.001\times\log{M_\mathrm{273}}+0.0057\times(\log{M_\mathrm{273}})^2$ with an uncertainty of 0.033.

We now proceed to compute the $\alpha$ parameter using both hydrodynamical as well as dark matter only simulations. 
For this purpose, we use the MaGICC \citep{brook12} and NIHAO \citep{Wang_15} hydrodynamic simulations, selecting Milky Way-like galaxies that have a similar circular velocity as the one of our Galaxy: we found that when V$_\mathrm{circ}$ at 19 kpc is consistent with observations, the obtained value of $\alpha$ agrees well between simulations and observations (NIHAO/MaGICC shown as green points in Fig.~\ref{fig:m_alpha}).

For a more statistically significant sample, we used the ELVIS dark matter only (DMO) simulation set \citep{Garrison-Kimmel_14}, which, being simulations of the local group of galaxies, have the advantage that the main halos have a similar environment as the MW, i.e. there is a M31 sized halo at about the correct distance from the MW and no other massive halo within 2.8 Mpc.  We selected all MW and M31 halos, for a total of 24 objects, and retrieved the masses at the bins borders (19, 64, 77 and 273 kpc) which allowed us to again compute $\alpha$  by interpolating between inner-most and outer-most mass using a power-law.
We compared these results with the $\alpha$ obtained from hydrodynamical simulations and from our observations in Fig.~\ref{fig:m_alpha}, where the 24 ELVIS points are cyan. As expected, the value of $\alpha$ in the DMO case is lower than in the hydrodynamical simulations and in the observed case, since baryons cool towards the centre of the halo making the profile more concentrated for r$\rightarrow$0.

In summary, in the mass range probed by the hydrodynamical simulations ($\sim 1.2 \times 10^{12}$M$_{\odot}$) the average values of $\alpha$ derived from both simulations and  MW observations fall in the range $\alpha \sim$0.3-0.4.

%%%%%%%%%%%%%%%%%%%%%%%%%%%%%

\subsection{Determination of the anisotropy parameter $\beta$}
\label{section:beta}

The velocity anisotropy is defined as 
\begin{equation} \label{eq:beta}
\beta=1-(\sigma_\theta^2+\sigma_\phi^2)/(2\,\sigma_\mathrm{r}^2) 
\end{equation}

\noindent where the terms are the velocity dispersions in spherical coordinates relative to the Galactic centre. Given the assumption that the velocity anisotropy is independent of radius for the Wa10 estimator, we first derive $\beta$ for the full sample. 

In order to calculate the velocity dispersion in the radial and tangential directions, we first transform the observed heliocentric systemic line-of-sight velocities and proper motions of each galaxy into velocities in a Galactocentric system; the associated uncertainties are estimated by Monte Carlo simulations, considering the uncertainties (statistical and systematic) in proper motions, line of sight velocities, distance of the galaxy, and distance (R$_0$) \citep{Bland_16,Abuter_19} and velocity of the Sun \citep{Reid_04,Schoenrich_10}  relative to the Galactic centre. Here, the most important contribution by far comes from the proper motion uncertainties.
The average and standard deviation of the velocities obtained from the Monte Carlo simulations are then used as the value and uncertainty of each galaxy's velocity.

For obtaining velocity dispersions, and therefore $\beta$, care is needed to ensure that large proper motion uncertainties are not interpreted as enhanced tangential dispersion, which would bias $\beta$ towards more negative values.
To this aim, we use a Bayesian approach very similar to that by \citet{Fritz_18b}. We run the MultiNest code \citep{Feroz_09, Buchner_14}, which is a multimodal nested sampling algorithm, to perform the posterior parameter estimation. We assume that the velocity distributions are Gaussian, taking into account measurement uncertainties on the individual quantities to determine the intrinsic spreads.
We treat each component of the velocity dispersion separately and fix the average velocity to zero. 
We adopted flat priors for the velocity dispersion components between 0 and 500 km/s.
The dispersions are determined well enough that we do not expect the choice of the prior to have a significant influence. 

The distribution of velocity anisotropies is derived by applying Eq.~\ref{eq:beta} to random draws of values from the posterior distribution of the velocity dispersion in the radial, azimuthal and polar direction.

\begin{figure}
    \includegraphics[width=0.70\columnwidth, angle=-90]{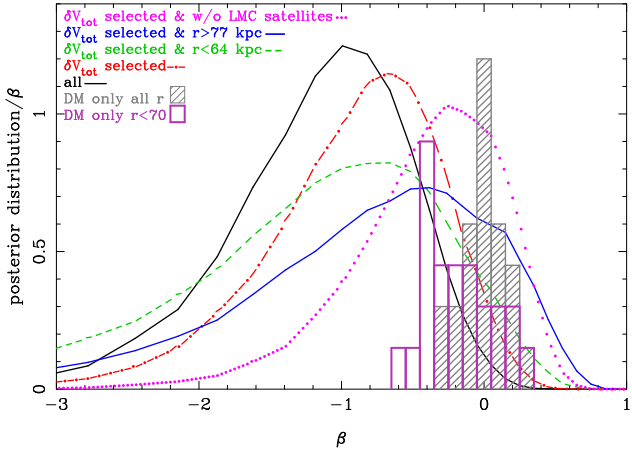}
 \caption{$\beta$ anisotropy parameter probability distributions, for different samples of satellites. 
 For the case excluding likely former LMC satellites we  only show the most extreme case, where all are excluded. Plotted is the binned probability scaled by the width of the bin. We also show the most likely $\beta$  from 24 DM only simulations from the ELVIS suite, both for all subhalos within 280 kpc, and once only the ones within 70 kpc are used. These two distributions are scaled such that they fit on the plot.
 } \label{fig:beta}
\end{figure}

Fig.~\ref{fig:beta} shows the $\beta$ obtained for the  different samples of tracers being considered.
The full sample of 45 satellites, whose posterior distribution is shown as solid black line in Fig.~\ref{fig:beta}, yields an anisotropy  $\beta=-0.92^{+0.49}_{-0.65}$ \footnote{The uncertainties represent the 68.3\% confidence interval around the median all through the paper.} very similar to the determination by \citet{Riley_18} ($\beta=-1.05^{+0.39}_{-0.49}$), which is not surprising since our sample has only 7 more galaxies compared to theirs\footnote{On the other hand, our uncertainties on the velocity anisotropy are larger than theirs, and we suspect that it is because we do not assume $\sigma_\theta=\sigma_\phi$; indeed the measurements from our sample suggest that they are significantly different from each other ($\sigma_\phi=101\pm13$ km/s and  $\sigma_\theta=192\pm24$ km/s).}. Allowing the mean velocities to be different from zero, $\beta$ changes slightly to $\beta=-0.61^{+0.39}_{-0.54}$, although its value is fully compatible with the previous determination within 1$\,\sigma$\footnote{While we obtain an average radial and azimuthal velocity consistent with zero, we find v$_\theta=-86\pm28$ km/s, which is not unexpected because it is known that the majority of galaxies rotate in one direction within the vast polar structure \citep{Pawlowski_12,Fritz_18a}.}.

 If we restrict the sample to the 36 satellite galaxies with the smallest uncertainties on the total velocity (the sample shown in blue in Fig.~\ref{fig:sample}), the velocity anisotropy becomes 
 $\beta=-0.67^{+0.45}_{-0.62}$, still consistent with the value from the full sample, but less tangential. This case is represented by a red posterior distribution in Fig.~\ref{fig:beta}. 
 
 It is likely that the assumption of a velocity anisotropy independent of radius made in the Wa10 mass estimator is violated in reality, as e.g. indicated by the data \citep{Riley_18} and as expected due to tidal disruption of satellites at small distances \citep{Kelley_18,Samuel_19}. Since this could have an impact on the inferred mass, here we test for such radial variations of $\beta$ (see also Section~\ref{sec:simulations}), by considering separately the objects belonging to the inner and outer samples defined in Section~\ref{section:data}. The inner sample yields $\beta=-0.76^{+0.63}_{-0.96}$ (shown as a green distribution in Fig.~\ref{fig:beta}), while
 for the outer one we get $\beta=-0.41^{+0.56}_{-0.90}$ (shown in blue in Fig.~\ref{fig:beta}).  While this might hint to a slight increase in the value of this parameter as a function of radius, the uncertainties are at present too large to pin this down. In any case, such a small difference in the values of $\beta$ would cause a negligible variation in the mass determination with the Wa10 estimator \citep[see Fig.~1, left panel, for the $\gamma=2$ case in][]{Dicintio_12}.

We also investigate whether the exclusion of the most likely, as well as tentative, former LMC satellites does have an effect on $\beta$, finding that the value becomes closer to isotropic ($\beta=-0.32^{+0.39}_{-0.53}$ in the former case and $\beta=-0.21^{+0.37}_{-0.51}$ in the latter case, see also Fig.~\ref{fig:beta}). This is somewhat expected, given the large tangential velocities of the satellites possibly associated to the LMC. Nonetheless, this variation is not expected to have a significant impact on the MW mass determination \citep{Dicintio_12}.

Since we will use cosmological simulations to address possible biases in the MW mass estimate (see Section~\ref{sec:simulations}) here we also compute the anisotropy parameter for the ELVIS simulations. Considering all the subhalos within 280 kpc from the Galactic centre, we obtain an average of $\beta=0.00$ with a scatter of 0.15 over all simulations (shown as grey histogram in Fig.~\ref{fig:beta}), 
while when  selecting only satellites within r$<70$ kpc,
 $\beta$\ is $-0.18$ (purple histogram in Fig.~\ref{fig:beta}).  
 
 Note that we have not included the determination of $\beta$ from hydrodynamical simulations since we found that they contain a lower number of luminous sub-haloes,  particularly in the inner regions, compared to observations, which makes the calculation less meaningful. This also applies to the calculation of $\gamma$ in the next section.

\subsection{Determination of tracer power-law index $\gamma$}
\label{section:gamma}

The parameter $\gamma$ is the power-law index of the tracer number density distribution. When considering the cumulative radial distribution of the satellites, $N_\mathrm{cum}$, this follows a power-law of index ${3-\gamma}$. 

The main difficulty in the determination of $\gamma$ originates in its dependence on the completeness of the galaxy surveys from which the satellites were detected.  \citet{Koposov_08} calculated that at 260\,kpc, approximately the distance of Leo~I, SDSS \citep{York_00} is complete down to an absolute magnitude M$_V=-5.9$. ATLAS \citep{Shanks_15}, Pan-STARRS-1 \citep{Chambers_16} ($\delta>-30$), DES \citep{Abbott_18} and some other smaller programs  \citep{Torrealba_18,Koposov_18,Nidever_17} have surveyed most of the remaining sky to similar or higher depths \citep{Jethwa_16}. Hence we make the assumption that these surveys are all complete at least to an absolute magnitude M$_V=-5.9$ (see recent quantification of the completeness of PS1 by \citet{Drlica-Wagner_19}).

 Already Pan-STARRS-1 and the NOAO all sky catalogue cover together nearly the full sky besides a few small regions. In practice, the main problem is not that some regions are missing completely, but that these surveys are less effective in detecting satellites closer to the Galactic plane due to the greater density of Galactic stars and extinction. 
 
 \begin{figure}
    \includegraphics[width=0.70\columnwidth, angle=-90]{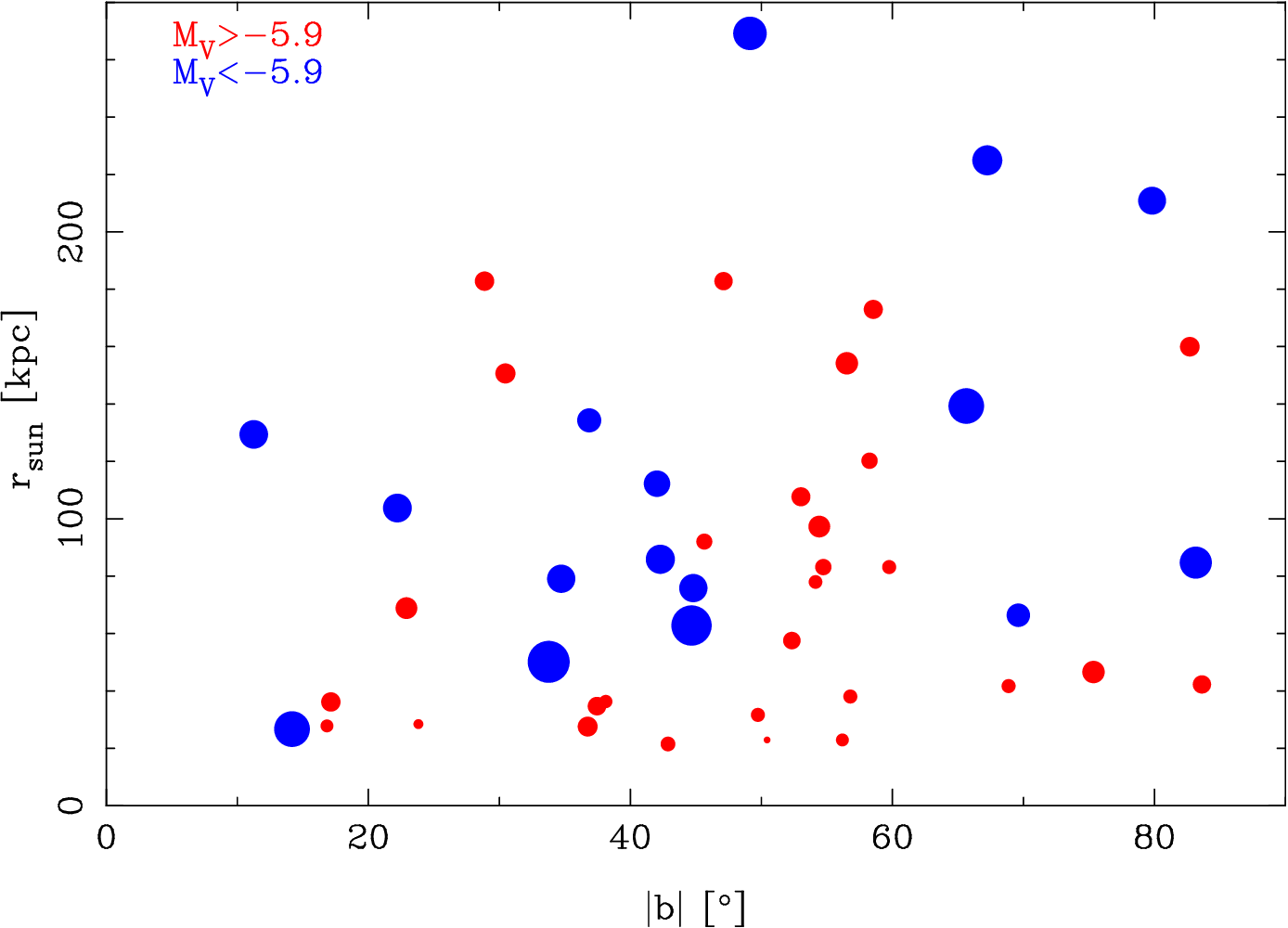}
        \includegraphics[width=0.70\columnwidth, angle=-90]{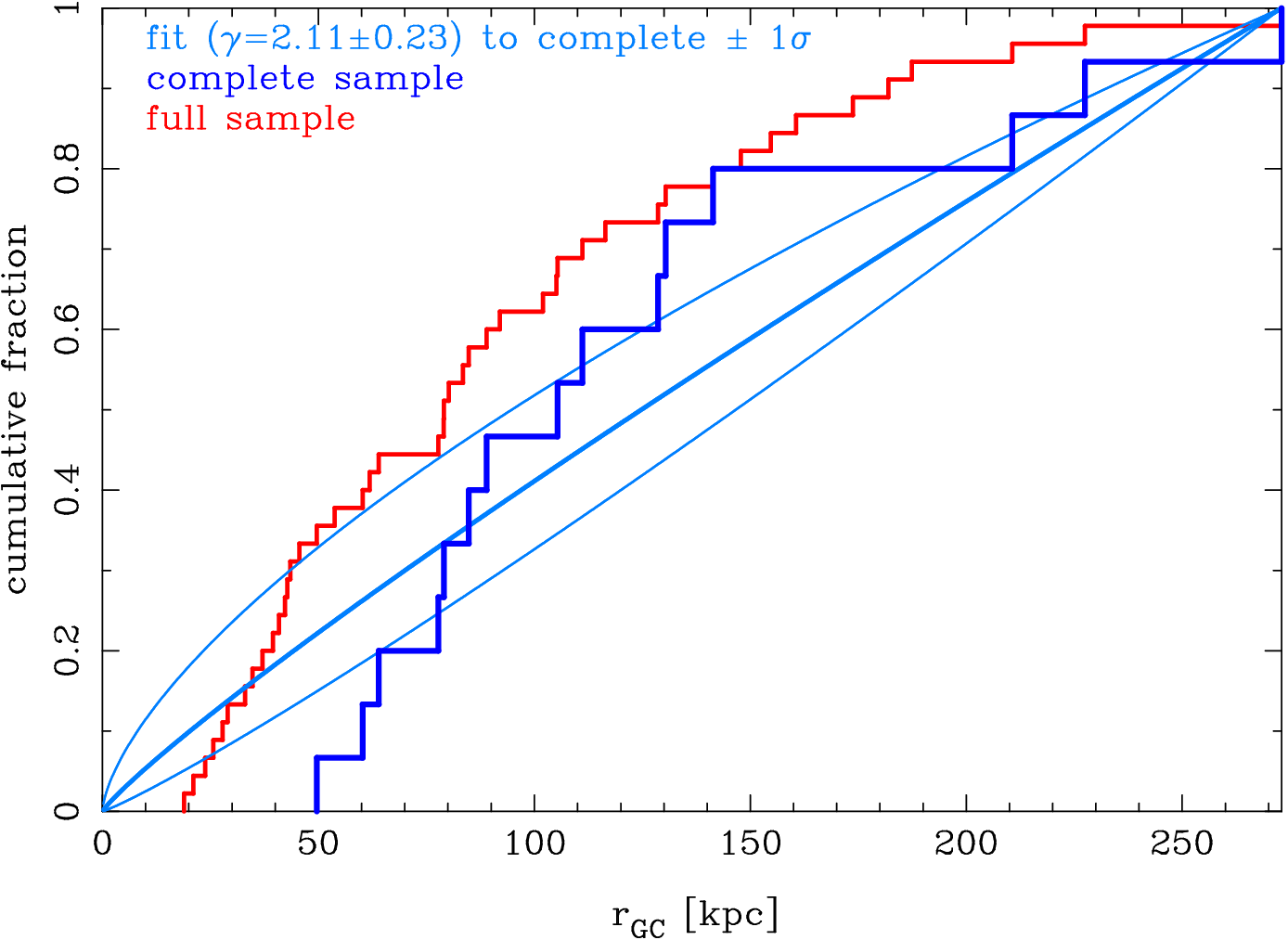}
 \caption{\textit{Top panel}: Heliocentric distance of the satellite galaxies against the absolute value of their galactic latitude.  Galaxies in the complete luminosity range \citep{Koposov_08} at high latitudes are coloured in blue, while the full sample is shown  in red. The size of the circles is larger for brighter galaxies. \textit{Bottom panel}: Cumulative distribution of the full and of the complete sample, with corresponding fit to the latter, shown as cyan lines with 1$\,\sigma$ interval.
 } \label{fig:lats}
\end{figure}

To get a feeling of how poor the completeness is close to the Galactic plane, especially for the sample complete in luminosity, in Fig.~\ref{fig:lats} we plot the heliocentric distance against the absolute Galactic latitude ($|b|$) of the satellites in our sample, where the size
of the circles is larger for brighter galaxies. The magnitude-complete sample, made of all galaxies brighter than M$_V=-5.9$, is indicated in blue, while the full sample is shown in red.

First, we notice that the lack of any galaxies within $|b|=$11 degrees is probably caused by incompleteness, but it is only marginally relevant for the sample's radial distribution: since all satellites are at a distance from the Galactic Centre (GC) that is more than double the distance between the GC and the sun, galaxies at all distances are nearly equally missed due to the zone of avoidance.
Second, the galaxy closest to the plane is Antlia II \citep{Torrealba_18b}: with M$_V=-9.0$ and a distance of 130 kpc it is fainter and more distant than the median galaxy in the complete sample, and is at a fairly average Galactic longitude (l$=-95$\degree). It is therefore  unlikely that satellites with less extreme properties are missing within this  distance. 

In contrast, the galaxy with the second smallest |b| is Sagr~I, the closest and one of the brightest satellites with M$_V=-13.5$ (the median luminosity is M$_V=-9.2$ in the complete sample):  
 with l$=5.6\degree$ it is in a region where galaxies are particularly difficult to discover due to the bulge.  We therefore  assume that a  more distant and fainter galaxy would not have been discovered at the location of Sagr~I, and so we exclude Sagr~I from the main sample, while we keep Antlia~II. 
The next galaxy brighter than the completeness limit of \citet{Koposov_08} is Carina~I at $|b|=22\degree$: since it is about at the median distance and luminosity of the complete sample we include it. At larger |b| than Carina~I there is likely no bias, since many fainter galaxies have been discovered there. 

We are left with a total of 15 galaxies brighter than M$_V=-5.9$, which constitutes our magnitude-complete sample, indicated as blue circles in Fig.~\ref{fig:lats}, top panel. 

We now proceed to fit a single
scale free power-law to the radial distribution of our full and magnitude-complete satellite samples\footnote{We note, however, that in reality there is a dearth of satellite galaxies within 18 kpc from the Milky Way's centre: in our full sample the number of galaxies per radial interval is approximately constant, with a value of 0.5 galaxies per kpc between 18 and 48 kpc; however, we observe no galaxy within 18 kpc. Thus, the single power-law as defined in Wa10 must be used with caution.}, starting from a radius of r$\rm_{GC}$=0. The fit is shown in cyan in the bottom panel of Fig.~\ref{fig:lats} for the magnitude complete sample: we obtain our main index of $\gamma=2.11\pm0.23$. We use Monte Carlo simulations to obtain the uncertainty, and we also check whether distance uncertainties matter, finding that they are irrelevant compared to Poisson uncertainties. 
 We also fit the same sample without the SMC, since that galaxy is likely not independent of the LMC (see  Section~\ref{sec:lmc}) and obtain $\gamma=2.08\pm0.25$. 
 When we include all galaxies, $\gamma$ only increases slightly to $2.24\pm0.12$, meaning that an index larger than 2.24 seems unlikely. 
 
 Given the minor differences from the various cases, for the mass estimator we use the $\gamma$ index and uncertainty derived from the magnitude complete sample (2.11$\pm$0.23). 

\begin{figure}

        \includegraphics[width=0.70\columnwidth, angle=-90]{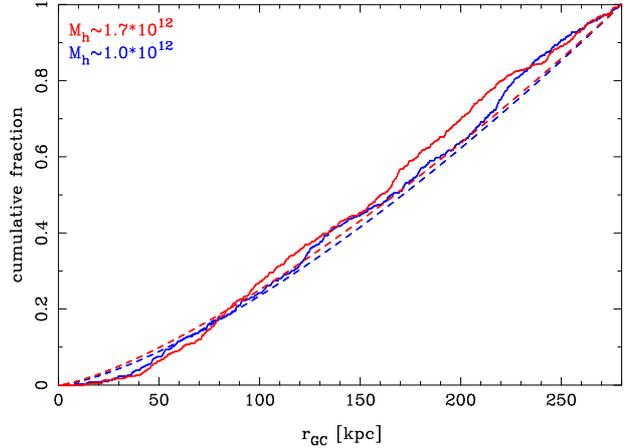}
 \caption{
 Cumulative number density profile with power-law fits for 2 typical ELVIS halos (in blue and red, with fits as dashed lines).  A single power-law is not a good fit to the central region of the subhalo distribution.} \label{fig:gamma}
\end{figure}

For comparison, and for the assessment of biases in the mass determination (see Section~\ref{sec:simulations}), we calculate the $\gamma$ index of the ELVIS simulations in the same way, using all subhalos within 273 kpc. We obtain $\gamma=1.60\pm0.09$ where the uncertainty comes from the scatter amongst the different simulated MW-halos. The slope is about 2$\,\sigma$ smaller than the measurement inferred for the MW. We also note that a power-law distribution does not appear to be a perfect representation of the radial distribution of sub-haloes in the ELVIS simulations: as it can be appreciated in Fig.~\ref{fig:gamma} for two examples, a power-law over-predicts the counts at small r and under-predicts them at intermediate and large radii.
 However, since the simulation uses only dark matter it is also not necessarily expected that they reproduce observations fully, due to effects like additional tidal destruction by the baryonic disk (see e.g. \citealt{Donghia_10}, \citealt{Garrison-Kimmel17}, \citealt{Riley_18}, \citealt{Kelley_18}). 

\section{Assessment of biases} 
\label{sec:simulations}

We use cosmological simulations to test whether issues such as selection effects and observational uncertainties might bias the mass estimates. For this, we focus on the ELVIS suite because of their statistical significance, since this sample consists of 24 Milky-Way-like halos, with a virial mass between 1.0 and $2.8\times10^{12}$ M$_\odot$. 

We select all resolved subhalos within 280 kpc, without requiring them to be bound, although in practice nearly all of them are \citep{Boylan-Kolchin_13}. The simulations contain between 324 and 813 (average of 539) subhalos. 

From each simulation we select a sample of 45 subhalos as close as possible in Galactocentric distance to our sample of galaxies. Overall the simulations contain enough subhalos to find a good match for all the observed galaxies, with the median difference in Galactocentric distance between best match and the corresponding observed galaxy of 0.2 kpc. The match is slightly worse at small radii ($r<30$ kpc), where the median difference in Galactocentric distance between best match and the corresponding observed galaxy is 1.4 kpc. The difference is noticeably larger for the subhalos of less massive hosts, reaching a median of 9 kpc at $r<30$ kpc in the worst case, due to the lower number of subhalos. 

To ensure that distance differences between simulations and observations do not matter (for example due to distance dependent tangential velocity uncertainties), we always adopt the values of the distance of the observed galaxies. We also adjust the position on the sky to the 'observed position', to preserve the influences of LOS velocities on the tangential velocities and other properties. However, we do not modify the velocity in spherical coordinates, so as to preserve the velocity anisotropy $\beta$. Since the rotation curve of the Milky Way is relatively flat in this radial range \citep{Iocco_15}, the impact of not adapting the velocities to the new distance is small. We checked for it and  find that the mass measured with the estimator changed at most by 0.9\% (on average by 0.1\%) for the full sample, and  at most by 4\% (on average by 0.5\%) for the sample within 64 kpc compared to the mass obtained using the original radii.

We use $\gamma$ and $\beta$ as determined from the full sample of sub-haloes, whose values have been presented in Sections~\ref{section:beta} and \ref{section:gamma}. 
In this way we can estimate the impact of the selection as independently as possible from the impact of uncertainties in $\beta$ and $\gamma$. For the mass index $\alpha$, we use relations between $\alpha$ and mass as derived in Section~\ref{section:alpha}.

\subsection{Observational uncertainty-free case}
\label{sec:sim-err-free}

 We first determine the correction factor of the Wa10 mass estimator when applied to simulations without including observational uncertainties.
For each of the three sampled regions (full, inner and outer radial range) we start with an initial value of $\alpha$ by using the average of all simulations, $\alpha=0.02$. We then iterate: we use the initial $\alpha$  in the first iteration to determine the mass and its correction factor (i.e. $f_\textrm{cor}$, the ratio between the estimated mass, mass$_{\rm est}$, and the true mass, mass$_{\rm true}$), and then use the previously derived mass dependency of $\alpha$ (Eq.~\ref{eq:mass_alpha_full} for the full range), to determine $\alpha$ for the second iteration. 
We then use this result for each mass to determine a new relation. A 3$^d$ and 4$^{th}$ iteration follow. By the 4$^{th}$ iteration the change in the correction factor is less than 0.1\%. 

\begin{figure}
    \includegraphics[width=0.70\columnwidth, angle=-90]{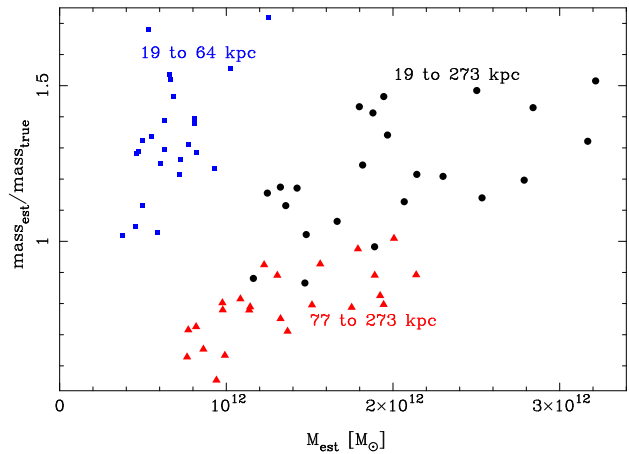}
 \caption{
 Test for biases in the Wa10 mass estimator:  ratio of mass estimated to true mass in the uncertainty free case for the dark matter only ELVIS cosmological simulations. The subhalos are selected such that they match the radial distribution of the MW satellites used. The results from the full sample are indicated in black, while for the inner and outer sample in  blue and red, respectively. Each symbol represents one MW-simulated galaxy in the ELVIS suite.
 } \label{fig:mas_sim}
\end{figure}

We show  in Fig.~\ref{fig:mas_sim} the ratio of estimated to known mass against the estimated mass for the different samples.
There is a clear offset, which is slightly larger for larger masses, particularly when the full sample is used.

 We calculate the significance of the trend, rescaling such that the reduced $\chi^2=1$:  
it follows that the trend has a significance of 3.4$\,\sigma$ for the full sample. 
On average, the correction factor is 1.23 for the full sample, 1.33 for the inner sample and 0.79 for the outer sample. Dividing the typical uncertainty through the correction factor gives a typical relative mass uncertainty of 0.16 for the full sample.

 We tested whether these biases are peculiar of the selected samples by repeating the analysis on samples of tracers in which the parameters ($\alpha$, $\beta$, $\gamma)$  fulfil  exactly the assumptions and definitions of the Wa10. These mock data-sets were constructed from distribution functions using Monte Carlo simulations\footnote{The mock data were kindly provided by Laura Watkins (private communication).}. We then select from these mock data `subhalos' at the same distances as the galaxies observed, and find that the mass obtained does not suffer from any bias. 
Therefore we are reassured that the mass estimator works in the ideal case, thus the reason for the bias must  be that the simulations do not fulfil all assumptions of the mass estimator.

 While overall the masses are overestimated, the  mass is underestimated in the outer sample. We suspect that the reason is that in the outer sample the majority of the subhalos are close to their apocenters. In contrast, in the inner part the majority are close to their pericenter, causing the opposite bias. The full sample is in between but also overestimates the mass, because  increasing incompleteness with  radius means that the subhalos are more likely close-in, and thus close to their pericenters.

\subsection{Adding observational uncertainties}
\label{sec:obs-errrors}

Observational uncertainties in the systemic proper motions typically bias velocities towards more tangential values, see e.g. \citet{Fritz_18a}. 
Since the mass estimator requires the same weighting for each galaxy, we cannot estimate the bias corrected measurement accounting for measurement uncertainties as in, for example, a Bayesian approach.   
Therefore, we apply forward modelling on the simulations and then estimate the correction factor as in Section~\ref{sec:sim-err-free}.

In practice, we only consider the effect of uncertainties in the proper motion estimates, as the other sources of error are negligible in comparison. This is implemented by extracting new proper motion values from a Gaussian distribution centred on the `observed' proper motions of the sub-haloes in the simulations and with dispersion equal to the proper motion uncertainty. The resulting properties are then propagated to the total velocities. For each simulated MW-analogue we perform 2000 independent draws, and calculate the average mass (and scatter) iteratively. 

\begin{figure}
    \includegraphics[width=0.70\columnwidth, angle=-90]{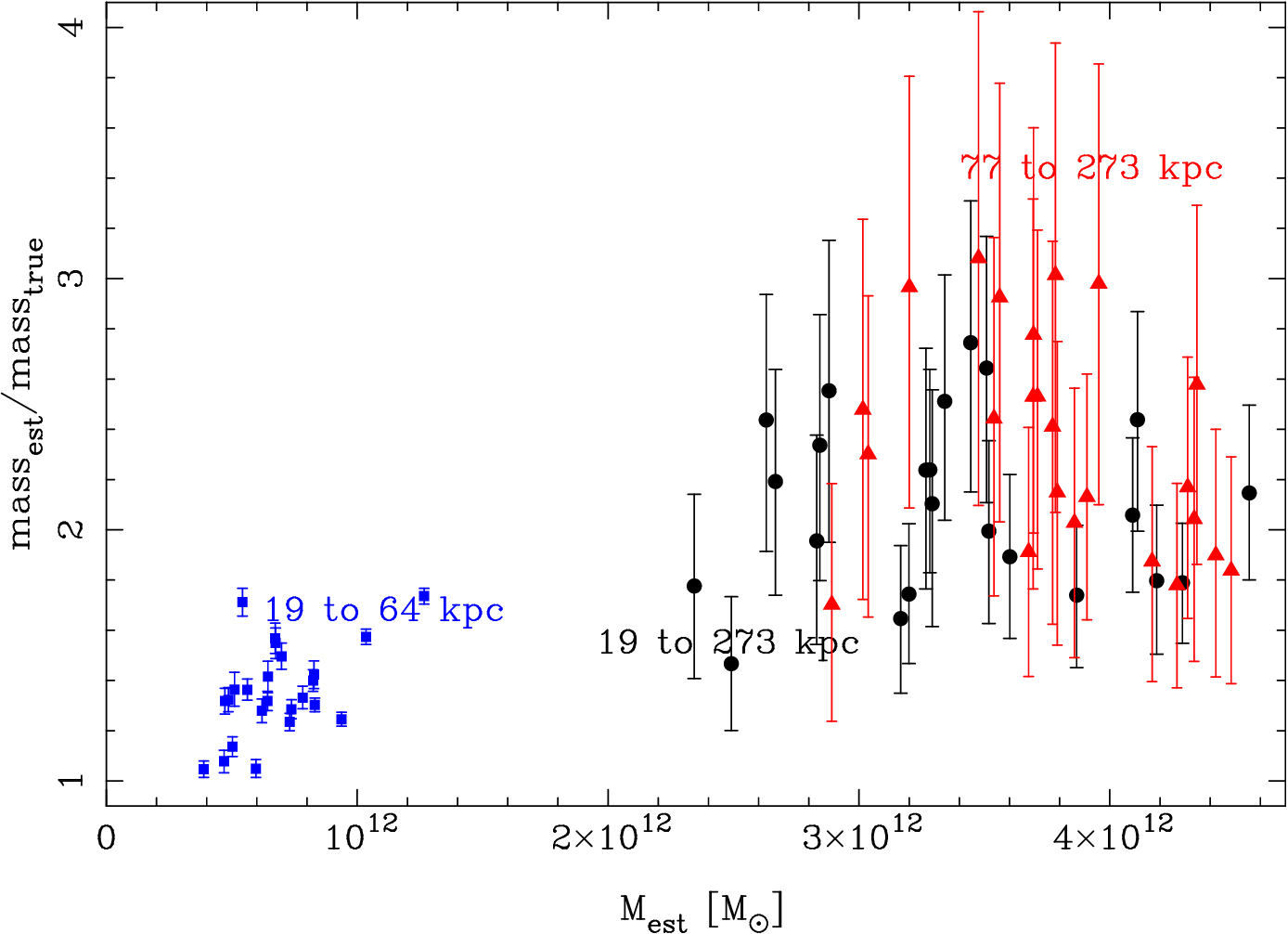}
    \includegraphics[width=0.70\columnwidth, angle=-90]{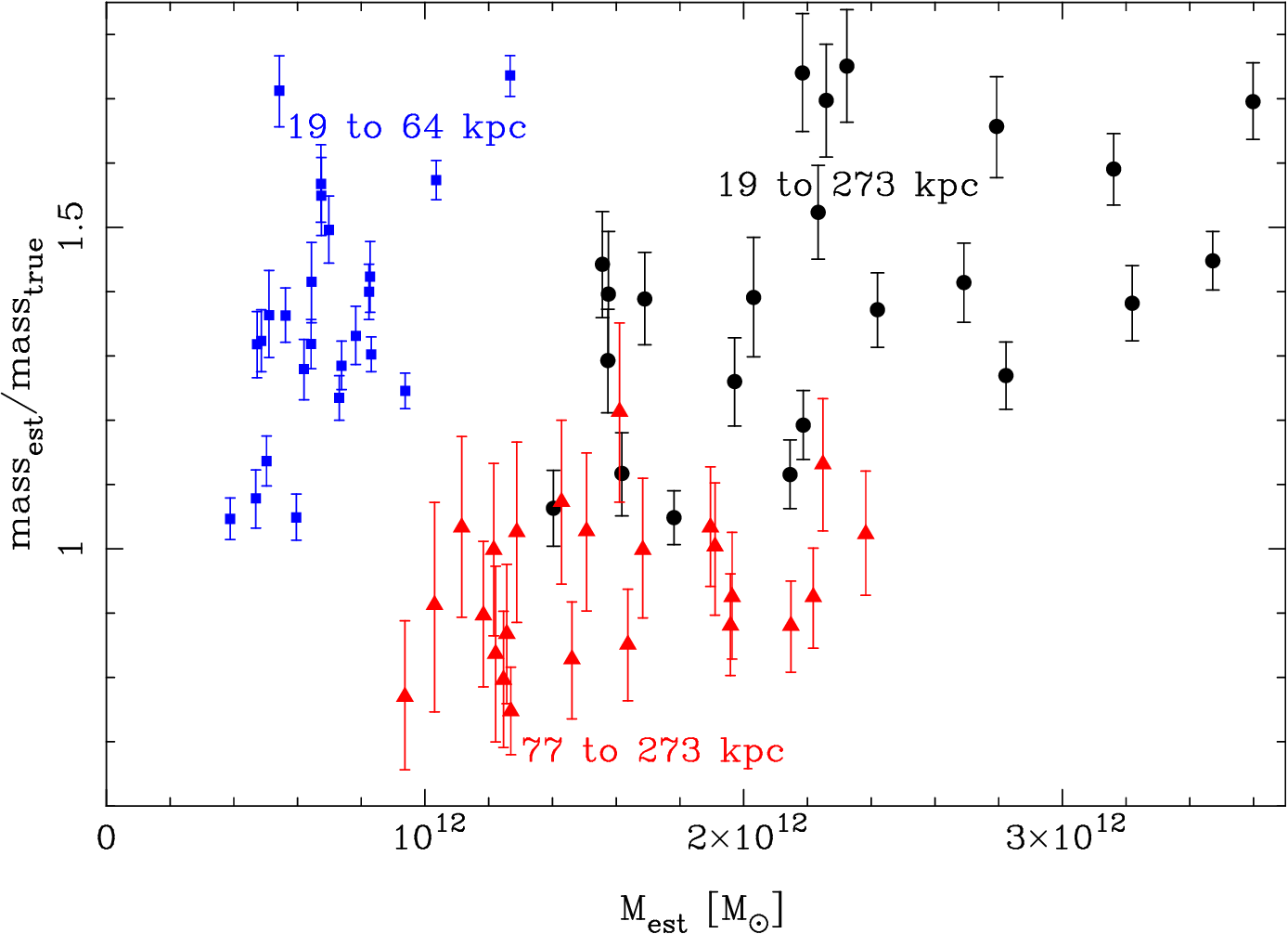}
\caption{Ratio of mass estimator to true mass in the dark matter only ELVIS simulations, with observational uncertainties added. Here, error-bars enclose the 1$\,\sigma$ intervals as obtained from 68.3\% confidence intervals. The subhalos are selected to match the radial distribution of the sample of MW satellites used in this analysis. This sample is then divided into an inner (blue), outer (red) and full (black) sample. Top panel: results for the whole sample. Bottom panel: we omit galaxies whose inclusion results in a loss of precision in the results, due to the large uncertainties in the velocity determination (see text). None of those is within 64 kpc, thus these points are identical in the two panels.
 } \label{fig:masser_bias}
\end{figure}

 As with the uncertainty free case, we obtain the correction factor as a function of the estimated mass (see Fig.~\ref{fig:masser_bias}). The obtained data are then fit as a linear function of $m_\mathrm{est}$, where we weight all simulations equally.  We also fit the scatter between the different trials as a linear function of $m_\mathrm{est}$ with equal weights to obtain the typical uncertainty.
 
 As shown in Fig.~\ref{fig:masser_bias} (top panel) the correction factor for the full sample increases from the uncertainty free case of 1.23 to 2.1. Dividing the uncertainty by the correction factor we obtain a typical relative uncertainty of 0.2 on the mass. When the scatter between the different simulations is added, the total relative mass uncertainty is  0.26 (to be compared to the value of 0.16 for the uncertainty free case). 

\begin{figure}
   \includegraphics[width=0.70\columnwidth, angle=-90]{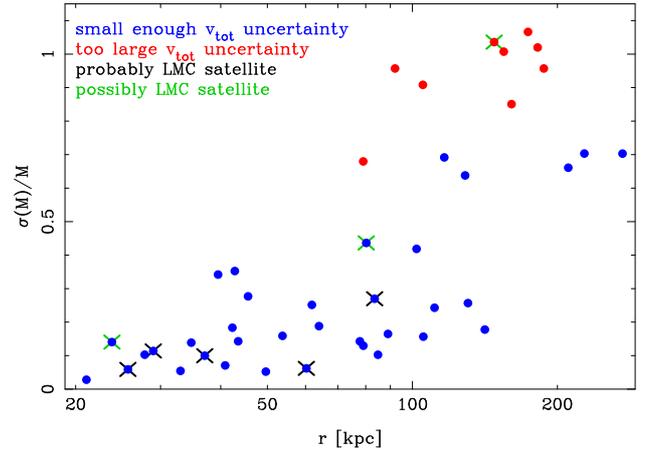}    
 \caption{Relative mass uncertainty due to proper motion uncertainties for each simulated subhalo matched to an observed satellite galaxy. The colour indicates the different samples. 
 } \label{fig:mas_sim_err}
\end{figure}

Can the exclusion of some objects reduce the relative mass uncertainty?  
To identify which objects might produce a large uncertainty in the mass estimate, we use the sample created from the simulations to obtain the relative mass uncertainty of each subhalo. Using the galaxy dependent part from Eq.~\ref{eq:wa10_orig}  we calculate the relative mass uncertainty as: 
\begin{equation}
   \frac{\sigma(M)}{M}=\frac{\sigma(v_\mathrm{tot}^2\,r_\mathrm{GC}^{0.02})}{ \langle v_\mathrm{tot}^2\,r_\mathrm{GC}^{0.02}\rangle}
\end{equation} where $\sigma$ is the 
standard deviation  obtained  using 2000 Monte Carlo draws, for each matched subhalo in a given ELVIS simulation. These values can be under- or over- estimated depending on the velocity that the subhalo happened to have in that given simulation. We then calculate the average $\sigma(M) /  M$ for each subhalo over the 24 simulations to remove this effect, such that we are left with the effect imprinted by the observational uncertainties. The results are shown in Fig.~\ref{fig:mas_sim_err}: as expected the uncertainties are larger for the most  distant galaxies, albeit with some scatter. 

The galaxies causing the largest  $\sigma$(M)/M are Leo~IV, Leo~V,  Pisces~II, Hydra~II, Columba~I, Aquarius~II, Reticulum~III, Canes Venatici~II. These correspond to a $\sigma(M)  /M \gtrsim 0.75$. Besides these, Horologium~II suffers from a large difference between the preferred proper motions derived by \citet{Fritz_18b} and \citet{Pace_19}. Excluding from the mass determination the sub-haloes that would be associated to these galaxies leads to the results shown in the lower panel of Fig.~\ref{fig:masser_bias}. The average correction factor is now smaller\footnote{The correction factor depends on mass, being usually smaller for smaller masses.}, typically 1.4,  and the average scatter over the different simulations is now only 0.07. That implies that the relative overall uncertainty is only 0.005 larger than without added uncertainties. An even stricter exclusion of galaxies is not advisable as it would lead to the exclusion of the most distant galaxies, which are especially useful to measure the mass within large radii.  

We also tested less strict cuts, but found that the obtained masses overlap well with the one obtained with our chosen cut. 

\subsection{Conclusions about biases}
\label{sec:con-biases}

We conclude that the biases in the mass estimators are likely due to deviations of the simulations from the assumptions of the mass estimator. The gravitational potential does not appear to be the culprit, as it is well approximated by a single power-law in the radial range considered. On the other hand, the radial density of the tracers, the subhalos, does not follow well a single power-law, as discussed in  Section~\ref{section:gamma}. For example, the simulations present a central deficit of sub-haloes with respect to the expectations from a single power-law fitted to the sub-haloes radial number density distribution. A reason for this deficit is disruption by tidal forces due low resolution in the simulations \citep{Vandenbosch_18a, Errani_19}. If that is the case, it is likely that the subhalos found at small distances are those close to their pericentre, and thus are biased towards larger velocities than assumed in the scale free Wa10 mass estimator. 
The behaviour of the sub-haloes velocity anisotropy, which is slightly tangential in the inner regions and radial further out (with a difference of $-0.18\pm0.04$ between the two), is likely  due to this effect. 

Since we are using the simulations to tackle the bias in the MW mass determination when applying the Wa10 estimator, and deduce a correction factor, it is important that simulations closely resemble the real systems.  On one hand, real satellite galaxies undergo baryonic effects that might enhance tidal disruption (see e.g. \citet{Donghia_10}, \citet{Kelley_18}, \citet{Samuel_19}), and that, by definition, are not included in dark-matter only simulations. 
On the other hand, a known issue with simulations is that subhalos are destroyed too rapidly, see \citet{Vandenbosch_18a, Errani_19}.  
Thus these two effects act in opposite directions. 
Observationally, the radial trend in $\beta$ hints to a more efficient destruction than in DM only simulations, while the high (concentrated) value of $\gamma$ hints to the opposite. 

We conclude that there is no clear evidence that we under- or overestimate the correction factor and thus we use the values calculated in Section~\ref{sec:obs-errrors}. It is possible that the estimate of the correction factor is sub-optimal, due to mismatches between the simulations and the true Milky Way: in this regard, it would be useful to revise this determination when hydrodynamic simulations will be able to provide a satisfactory match to the population of Milky Way satellite galaxies. 

%%%%%%%%%%%%%%%%%%%%%%%%%%%%%%%%%%%%
\section{Assumptions regarding satellites of the LMC} 
\label{sec:lmc}
%%%%%%%%%%%%%%%%%%%%%%%%%%%%%%%%%%%%

In this section we lay the ground for testing the effect that  the LMC has on the MW mass estimate, in two respects: accounting for the fact that some of the galaxies in the sample might be former LMC satellites; and taking into consideration the gravitational pull that a massive LMC is expected to exert on the MW \citep{Gomez_15}. 

As a first step, we determine how the mass changes when excluding galaxies that are likely to be former satellites of the LMC. These systems share similar orbits as their former host, and thus their inclusion would lead to an underestimated uncertainty if they are assumed to be independent tracers. Since we use cosmological simulations to estimate the uncertainty on the MW mass estimate, the effect of group infall is in principle included; therefore, we do not exclude small (and uncertain) groups such as the potential Crater-Leo~II group from our sample \citep[e.g.][]{Fritz_18a}. The LMC is, however, an unusually large satellite (satellites as massive as the LMC are rare in observations and simulations, e.g. \citealt{Liu_11}, \citealt{Boylan-Kolchin_10}). We therefore exclude its satellites in this test, but we do include the LMC itself, since one tracer should be used for a group.

The membership in the LMC associated group is still uncertain for galaxies with large proper motion uncertainties, see \citet{Fritz_18b}, which also implies that these systems are among those that are not included in our final sample.
 On the other hand, even when a proper motion is well known, other factors such as the mass of the LMC itself do play a role \citep{Kallivayalil_18}. 
 
 We classify the SMC,  Hyi~I, Car~II, Car~III and Horo~I as "probable" LMC group members \citep{Kallivayalil_18}, and Dra~II, Hydra~II \citep{Kallivayalil_18}, Phoenix~II as "possible" members \citep{Fritz_18b}. 
The possible members do not agree with the properties of the expected debris in \citet{Kallivayalil_18}, but this may  be due to differences between the LMC-analogue used in that work and the real LMC.  There are more potential members in the view of \citet{Pardy_19}, such as Fornax and Carina~I, but since these galaxies are now close to apocentre (while the LMC is at pericentre) we think that an association is unlikely (see also \citealt{Sales_11}). While they share the orbital plane of the LMC\footnote{They are members of the vast polar structure \citep{Pawlowski_12}.}, they do not have the other properties that would classify them as former satellites.

We end with three different samples: 1. all satellites, 2. a sample that excludes probable former LMC satellites and 3. a sample that excludes probable and possible former LMC satellites ("excludes all former LMC satellites").

To include the effects of the LMC gravitational pull on the MW, it would be ideal to have many analogues of LMC$+$MW in simulations, which is not yet possible. As an approximation we calculate the distance and total velocities of the satellite galaxies compared to the barycentre of MW and LMC. We tried several options for the relative LMC mass, but to give an idea of the potential size of the effect we use the most extreme plausible option, i.e. that the LMC has a fourth of the mass of the MW. That is slightly larger than recent estimates but not ruled out:  \citet{Penarrubia_16} obtained a mass ratio of $0.19\pm0.05$ from mostly LOS motions and timing arguments; the number of former LMC satellites points to a ratio of  $0.18\pm0.09$ \citep{Fritz_18b};  and the Orphan stream points to a mass ratio of between  0.12 to 0.19 \citep{Erkal_18}. 

Galaxies close to the Milky Way can adapt to the changing velocity of the MW, with the transition being at about 30 kpc \citep{Erkal_18}. Thus, galaxies further in should follow the Milky Way, even when considering the barycenter of the LMC+MW system. Since the transition is not sharp, we define a parameter for the strength of the reflex motion ($c_\mathrm{refl}$) calculated as: $$ c_\mathrm{refl}= (\tanh{(r_\mathrm{GC} \mathrm{[kpc]}-30)/10}+1)/2$$

\noindent When $c_\mathrm{refl}=0$  the velocity and position of a satellite is relative to the Milky Way, when  it  is  1  the  velocity  and  position  of  a  galaxy  is  relative to the barycentre of the LMC+MW system. 
It is larger than 0.5 for most galaxies. For the closest galaxy Sagr~I, it is 0.25. Therefore, the difference compared to using barycentric for all is small.

Since this correction is only approximate, we provide the MW mass estimate obtained in this way in addition to the case of referring positions and velocities of the satellite galaxies only to the MW. 

We combine these two cases with the above outlined three cases regarding former LMC satellites. Thus, we have in total six different combinations, see Table~\ref{KapSou2}.

%%%%%%%%%%%%%%%%%%%%%%%%%%%%%%%%%%%%
\section{The Mass of the Milky Way} 
\label{sec:MW-mass}

%%%%%%%%%%%%%%%%%%%%%%%%%%%%%%%%%%%%

In this section we obtain the mass of the MW and compare with previous results, as well as comparing the mass we obtain as a function of radius with simulations. 

\subsection{Derivation of the Milky Way mass} 
\label{sec:MW-mass_exp}

We now apply the Wa10 mass estimator to obtain the mass of the Milky Way, by using Eq.~\ref{eq:wa10_orig}. We adjust $M_\mathrm{<=out}$ by the correction factor ($f_\textrm{cor}$) to the correct mass $M_\mathrm{<=out\,cor}$:

\begin{equation}
M_\mathrm{<=out\,cor}=\frac{M_\mathrm{<=out}}{f_\mathrm{cor}}=\frac{1}{G\,f_\mathrm{cor}} \frac{\alpha+\gamma-2\beta}{3-2\beta}r_\mathrm{GC\,out}^{1-\alpha}  \langle v_\mathrm{tot}^2r_\mathrm{GC}^\alpha \rangle.
\label{eq:wa10_used}
\end{equation}

\noindent We use the distributions for $\beta$ and $\gamma$ obtained in Sections~\ref{section:beta} and \ref{section:gamma}. The mass is then obtained in an iterative process, since both the correction factor and the parameter $\alpha$ depend on the mass itself.  First we iterate until $\alpha$ converges, then add the mass independent uncertainty on $\alpha$ of  0.0706 (see Section~\ref{section:alpha})
as a Gaussian.
The number of iterations is set to be  high enough to ensure that the mass changes by less than 0.5\% at the end of the final iteration. 
The obtained mass is then scaled with the correction factor obtained in Section~\ref{sec:obs-errrors} to get the mass at the outermost radius. 
After five iterations we have a distribution of mass values that depend on $\alpha$, $\beta$ and $\gamma$. 
We then add the uncertainty of the correction factor obtained in Section~\ref{sec:obs-errrors} as a Gaussian  uncertainty. 
 We repeat this process for the different samples, i.e. those with and without LMC associated satellites,  as outlined in Section~\ref{sec:lmc}.

\begin{figure*}
    \includegraphics[width=0.70\columnwidth, angle=-90]{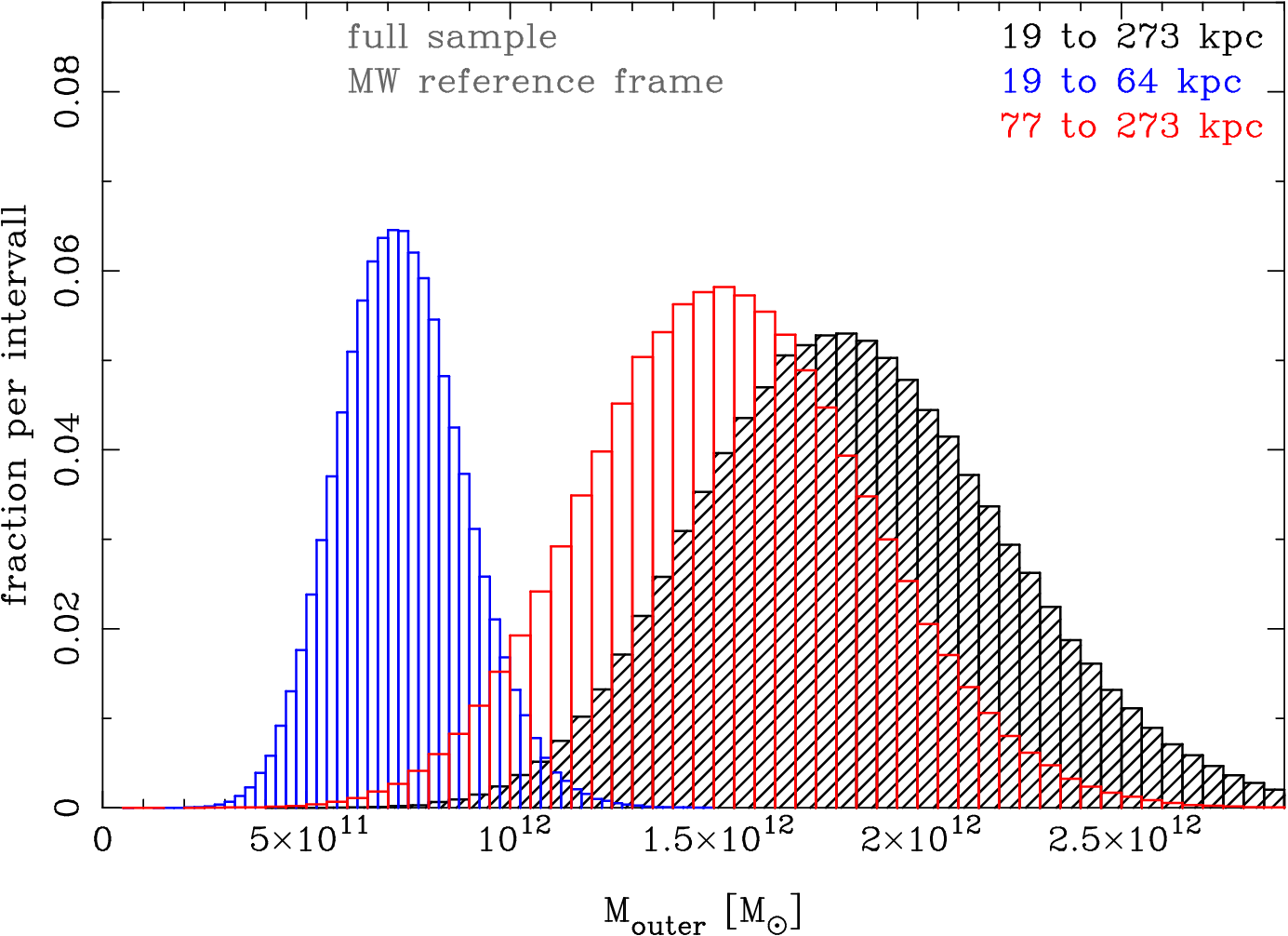}
        \includegraphics[width=0.70\columnwidth, angle=-90]{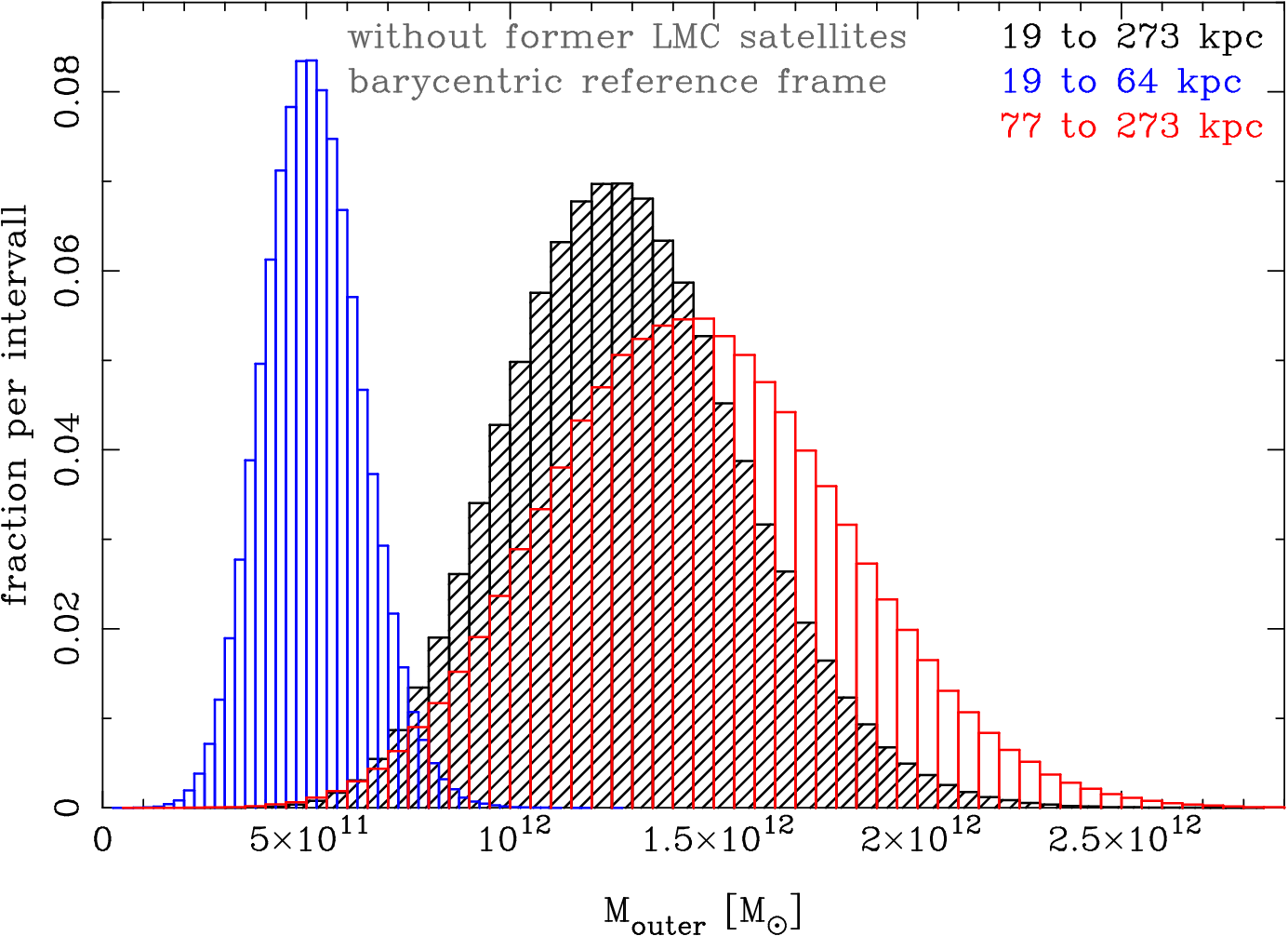}   
    \caption{Probability histograms of the Milky Way mass calculated within the inner, outer, and full radial ranges, shown in blue, red and black, respectively. 
    In the left panel we show  results for our full sample of satellite galaxies, for which we obtained a value of the MW mass within 273 kpc of   $M_\mathrm{273}=1.84^{+0.40}_{-0.36}\times10^{12}$M$_\odot$, while in the right panel we show results for the most restrictive sample in which all possible former satellites of the LMC are excluded, and for which the derived MW mass reads $M_\mathrm{273}=1.28^{+0.30}_{-0.27}\times10^{12}$M$_\odot$.
    In this last case, the velocities and distances of the tracers are relative to the centre of mass of the MW-LMC system. We show the two extreme estimates of mass: other cases provide values of mass in between these two.
    } 
\label{fig:gal-exclusion}
\end{figure*}

The values obtained for the MW mass for the different combination of cases are summarised in Table~\ref{KapSou2}. In all cases we calculated masses from the inner (19 to 64 kpc), outer (77 to 273 kpc), and full (19 to 273 kpc) radial ranges, showing the mass within the distance of the outermost satellite. Fig.~\ref{fig:gal-exclusion}  shows the distribution of masses for the most extreme cases: the mass derived from the full sample, centred on the MW (left panel); and from the most reduced sample,  relative to the centre of mass of the MW-LMC system, and disregarding all former LMC satellites (right panel). 
 
      \begin{table*}
      \caption[]{Mass estimates for the various cases. Col. 2 lists the selection of galaxies, i.e. whether former LMC satellites are excluded; col. 3 the reference frame, either MW only or the barycentre of MW and LMC; col. 4 gives the number of galaxies in the full, inner and outer sample and col. 5, 6, 7 give the median and the range of the 68.3\% confidence interval within X kpc derived from the full sample, inner sample and outer sample.    }
         \label{KapSou2}
      $
         \begin{array}{p{0.03\linewidth}lllllll}
         
            \hline
            case  & \mathrm{galaxy~selection} & \mathrm{reference~frame} & \mathrm{number~of~galaxies} & M_\mathrm{273} [M_\odot] & M_\mathrm{64}  [M_\odot] & M_\mathrm{outer,273}  [M_\odot]\\
            \hline
            
1 & \mathrm{full~sample} & \mathrm{MW} & 36/20/16 &    1.84_{-0.36}^{+0.40} \times 10^{12} & 0.73_{-0.15}^{+0.16} \times 10^{12} & 1.53_{-0.33}^{+0.35} \times 10^{12} \\
2 & \mathrm{full~sample} & \mathrm{barycentric} & 36/20/16 &  1.53_{-0.31}^{+0.34} \times 10^{12} & 0.63_{-0.12}^{+0.14} \times 10^{12} & 1.46_{-0.32}^{+0.33} \times 10^{12} \\
3 & \mathrm{without~probable~former~LMC~satellites} & \mathrm{MW} & 31/16/15 & 1.54_{-0.33}^{+0.37} \times 10^{12} & 0.62_{-0.14}^{+0.15} \times 10^{12} & 1.49_{-0.34}^{+0.36} \times 10^{12} \\
4 & \mathrm{without~probable~former~LMC~satellites} & \mathrm{barycentric} & 31/16/15 & 1.36_{-0.3}^{+0.33} \times 10^{12} & 0.54_{-0.12}^{+0.13} \times 10^{12}& 1.48_{-0.34}^{+0.36} \times 10^{12}\\ 
5 & \mathrm{without~all~former~LMC satellites} & \mathrm{MW} & 29/15/14 &  1.43_{-0.31}^{+0.34} \times 10^{12} & 0.6_{-0.14}^{+0.15} \times 10^{12}& 1.37_{-0.33}^{+0.35} \times 10^{12} \\
6 & \mathrm{without~all~former~LMC satellites} & \mathrm{barycentric} & 29/15/14 & 1.28_{-0.27}^{+0.3} \times 10^{12} & 0.51_{-0.11}^{+0.12} \times 10^{12} & 1.45_{-0.35}^{+0.38} \times 10^{12}\\
            \noalign{\smallskip}
            \hline
         \end{array}
     $
   \end{table*}

When all the tracer galaxies are used, we obtain the largest value of the mass  within 273 kpc (full radial range),  $M_\mathrm{273}=1.84^{+0.40}_{-0.36}\times10^{12}$M$_\odot$, while by excluding all potential LMC satellites, and computing the positions and velocities  relatively to the centre of mass of the MW-LMC system, the obtained mass is the smallest possible  with $M_\mathrm{273}=1.28^{+0.30}_{-0.27}\times10^{12}$M$_\odot$.\footnote{We note that here as in the rest of the paper $_\mathrm{X}$ stands for the full mass within X (e.g. 273) kpc.}
While it is unlikely that including all the galaxies centred on the MW is correct, since it is known that the LMC is on an eccentric orbit which biases the estimates towards large values, it is difficult to assess which of the other five cases is preferable: we therefore average them (see Table~\ref{KapSou2}) and add to the uncertainty the scatter over the median masses.    
 The result is $M_\mathrm{273}=1.43^{+0.35}_{-0.32}\times10^{12}$M$_\odot$, which is our preferred value for the MW mass within 273 kpc.

 The mass derived within the  outer radial range depends less on the choice of the sample, since there are less former LMC satellites in the outskirts of the MW: the outer mass varies only between 1.37 and 1.53 
 $\times10^{12}$ M$_\odot$, with an average of $M_\mathrm{273}=1.45^{+0.37}_{-0.35}\times10^{12}$M$_\odot$. 
 The two different estimates of the MW mass at 273 kpc agree well, although the correction factor is different and the sample differs by half.
 Within 64 kpc, instead, the mass varies from 0.51 to 0.73  $\times10^{12}$ M$_\odot$, 
  with an average of  $M_\mathrm{64}=0.58^{+0.15}_{-0.14}\times10^{12}$M$_\odot$.

Finally, we convert the mass within 273 kpc to the virial mass. 
Since our sample nearly extends to the virial radius, a large extrapolation is not required. We tried several methods. Firstly by using the virial mass and mass profile given by ELVIS simulations. We also used a typical NFW profile which concentration-mass relation from Planck cosmology and overdensity of 97. We then used an NFW profile with a  value of concentration which is twice the dark matter only value, since the contraction in Milky Way-like hydrodynamic simulations is found to be higher than in dark matter only simulations at this mass (see Fig.~4 in \citealt{DiCintio_14}).

The difference between the largest and smaller estimates of the virial mass is 6\%\footnote{Our definition of the virial mass assumes $\Delta=97$. Other 'virial' masses are indicated when used in comparisons, for example $M_\mathrm{vir,200}$. The virial masses include the full mass, not only DM.}, significantly less than our overall uncertainty. We use the median value, which is obtained with the NFW extrapolation and twice the standard concentration. 
\textit{This results in  $M_\mathrm{vir}=1.51^{+0.45}_{-0.40} \times 10^{12}M_{\odot}$, which is our preferred value for the virial mass of the Milky Way, corresponding to a virial radius of R$_\mathrm{vir}=308\pm29$ kpc.}
 
We then explore which of the parameters of the mass estimator most affects the uncertainty on the mass.
We vary the different power-law indexes separately for the full sample, setting the fixed indices to their median values. 
We obtain the following relative uncertainties: for $\alpha$ 8.1\%; for $\beta$ 4.3\%; for $\gamma$ 4.4\%. 
Thus, uncertainty in $\alpha$, the galaxy mass profile, is what matters the most, while the uncertainty in the tracer profile ($\gamma$) and the anisotropy parameter ($\beta$) are of similar, lower, importance. Therefore, although the $\beta$ uncertainty is large, its impact is relatively minor because we use full velocities and not only radial (approximately line-of-sight) or tangential (proper motions) velocities. This is in agreement with the analytic derivation, shown in \citet{Dicintio_12}, of the relative uncertainties  of each parameter and their impact on the mass estimators uncertainty.

The uncertainties on these three free parameters all together contribute an uncertainty of 13.5\%, which is more than expected from simply adding the uncertainties individually, since such parameters are correlated: a change in $\beta$ and $\gamma$ modifies the expected mass which in turn changes $\alpha$, that depends on the mass. 
Thus, their combined contribution is nearly the same as the
uncertainty caused by the calibration of the mass estimator itself, which is 15\%. 
Improved determinations of $\alpha$, $\beta$ and $\gamma$ would therefore result in significant improvement in the  application of the Wa10 mass estimator.

\begin{figure*}
        \includegraphics[width=1.49\columnwidth, angle=-90]{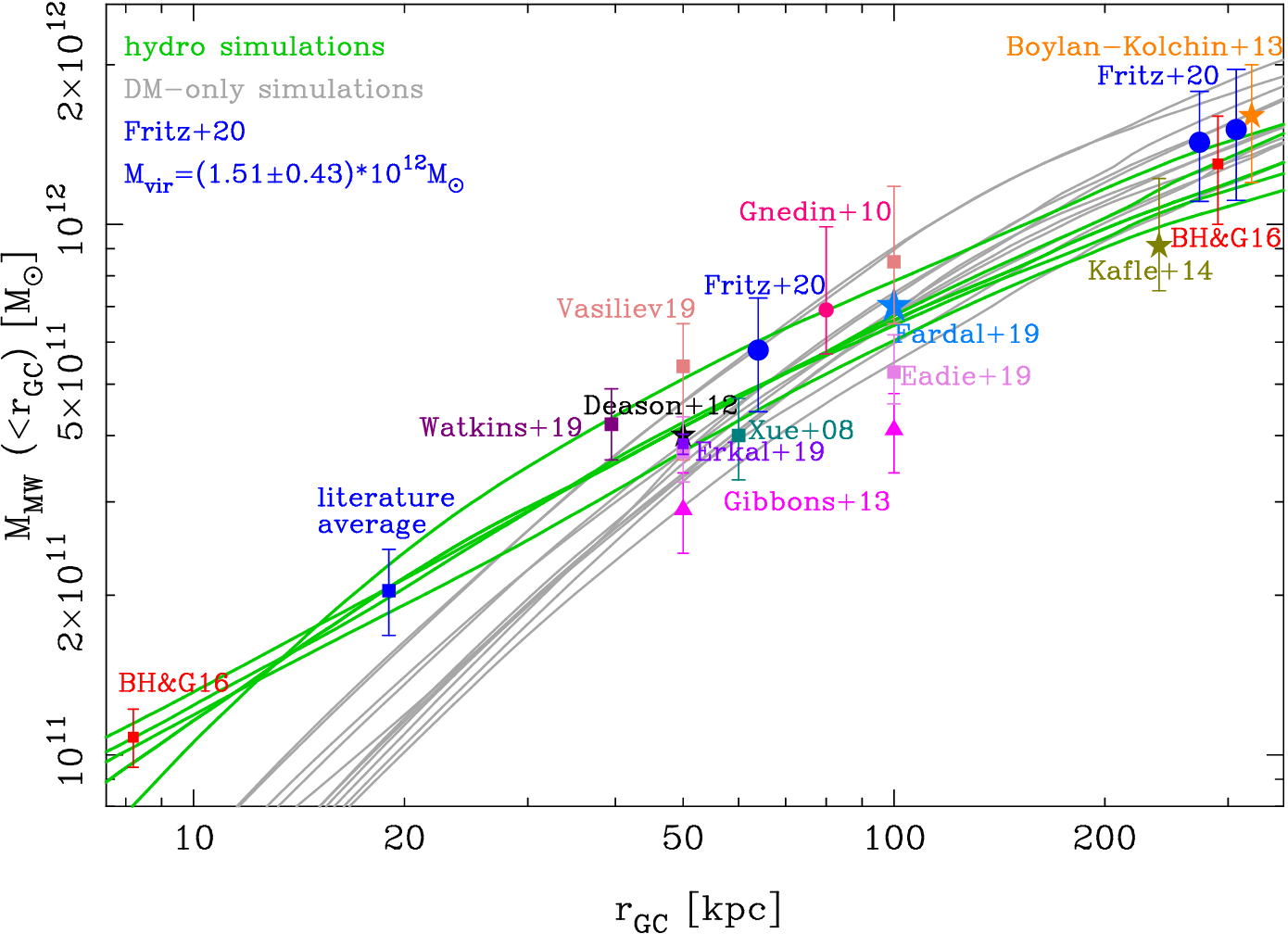} 
    \caption{Radial distribution of the mass of the Milky Way, in Galactocentric coordinates. We show with blue dots our measurements at 64, 273 and 293 kpc, where the outer most point is the virial radius. 1$\,\sigma$ uncertainties are also shown.
    The blue box at 19 kpc is our average estimate from \citet{Kuepper_15}, \citet{Bovy_16}, \citet{Watkins_18}.
    With grey lines we show the mass profiles from  the DM only ELVIS simulations, which match within 1$\,\sigma$ our mass at 273 kpc.  With green lines we show the total mass from  NIHAO and MaGICC hydrodynamical simulations, which match within  1$\,\sigma$ the mass profile at 19 kpc. The remaining points are previous literature results \citep{Vasiliev_19,Eadie_19,Watkins_18,Kafle_14,Deason_12,Fardal_19,Xue_08,Gibbons_14,Bland_16,Boylan-Kolchin_13}, based on different methodologies (see text for more detail). The estimates of \citet{Boylan-Kolchin_13} and 
    \citet{Kafle_14} are two of the most extreme virial mass estimates, see text for more values in between.
    }
\label{fig:mass-rad}
\end{figure*}

\subsection{Comparison with simulations}
\label{sec:mass_comp_sim_obs}

We now compare our estimates of mass at different radii with the MW mass profile coming from cosmological simulations. 

From the ELVIS suite we select DM-only halos whose mass at 273 kpc is within  1$-\sigma$ interval of our mass estimate, as shown in Fig.~\ref{fig:mass-rad}: the virial mass of such halos range between 1.06 and 1.90 $\times10^{12}$M$_\odot$, and they follow an NFW profile.

At 64 kpc, instead, our measurement of the MW mass lies above the average of the DM-only predictions: that is expected because DM-only halos neglect the contribution of baryons. This results in an even clearer disagreement at 19 kpc and further in, where the contribution of baryons is increasingly important. 

We therefore compare to hydrodynamical simulations, for which we use the NIHAO and MaGICC simulations, considering only those simulated MWs within  1$\,\sigma$ of the  MW' observed $v_\mathrm{circ}$ value at 19 kpc. 
This gives us five simulations, that span a virial mass range between 1.06 and 1.40$\times10^{12}$ M$_\odot$. 
We see that further out than 19 kpc most of these hydro simulations are below our most likely estimate, however 4 out of 5 still lie within our uncertainty at 273 kpc. 

Over-all, the mass profile derived from our measurements agrees well with recent hydro-dynamical simulations of the MW, despite of the fact that we did not use for our mass estimates. This shows that the simplifications present in the simulations and the assumptions made for our mass estimates probably do not have a large impact on the mass estimate.

\subsection{Comparison with previous results}

Fig.~\ref{fig:mass-rad} shows the comparison of our measurement of the MW mass with those from the literature.
We concentrate on those measurements over a similar radial range as our work (r$>$30 kpc), that do not use large extrapolations  and/or that do not combine different types of tracers that might follow a different radial distribution like \citet{Battaglia_05}, \citet{Dehnen_06}, \citet{Karukes_19b} and \citet{Cautun_19}. 
 We list the measurements sorted by the tracer used,  in order to explore whether the choice of tracer matters for the derivation of the mass. 

We first note that our determination agrees well with  the virial mass of  $M_\mathrm{vir}=1.3\pm0.3\times10^{12}$M$_\odot$, shown as a red square in Fig.~\ref{fig:mass-rad}, as estimated in the review of \citet{Bland_16}, using an average of halo stellar kinematic measurements (see also \citealt{Wang_15a} and \citealt{Wang_19} for other reviews). 

The advantage of using halo stars for this type of determinations is that they are more numerous than other tracers, such as globular clusters, satellite galaxies or streams. \citet{Xue_08} obtain from the SDSS BHB stars a mass at 60 kpc of $M_\mathrm{60}=0.4\pm0.07\times10^{12}$M$_\odot$. \citet{Deason_12} obtain from SDSS BHB stars a value of $M_\mathrm{50}=0.4\times10^{12}$M$_\odot$.
\citet{Gnedin_10} uses also BHB stars from a dedicated survey to measure $M_\mathrm{80}=0.69^{+0.30}_{-0.12}\times10^{12}$M$_\odot$. \citet{Kafle_14} use K-giant stars, in addition to BHB ones, out to 155 kpc, to obtain $M_\mathrm{vir}=0.91^{+0.31}_{-0.16}\times10^{12}$M$_\odot$, corresponding to a R$_\mathrm{vir}$ of $\sim$239 kpc, shown as a green star in Fig.~\ref{fig:mass-rad}. We note that this value is based on R$_0=8.5$ kpc,  and that reducing R$_0$, as suggested by \citet{Bland_16} and \citet{Abuter_19}, would increase the value of the mass.
 \citet{Huang_16} obtain $M_\mathrm{vir}=0.97^{+0.07}_{-0.08}\times10^{12}$M$_\odot$, mainly from line-of-sight velocities of giant stars out to distances of 100 kpc,

The escape speed of local stars can also constrain the virial mass. Using this method,  \citet{Piffl_14} obtain $M_\mathrm{vir,200}=1.6^{+0.5}_{-0.4}\times10^{12}$M$_\odot$ using the RAVE survey. 
\citet{Monari_18} use a similar method based on counter rotating stars from \textit{Gaia} DR2 to obtain $M_\mathrm{vir,200}=1.28^{+0.68}_{-0.5}\times10^{12}$M$_\odot$. Similarly, \citet{Hattori_18} use high velocity stars from \textit{Gaia} DR2 to obtain  $M_\mathrm{vir,200}\approx1.4\times10^{12}$M$_\odot$.
 \citet{Deason_19} also use high velocity stars from
\textit{Gaia} DR2 to obtain $M_\mathrm{vir,200}=1.00^{+0.31}_{-0.24}\times10^{12}$M$_\odot$.  
When better accounting for biases, especially  those affecting the halos that are dominated by early mergers, by using simulations  from \citet{Grand_19},  \citet{Deason_19}  get an increased value of $M_\mathrm{vir,200}=1.29^{+0.37}_{-0.47}\times$M$_\odot$. 
We note that since all these five virial masses are at an overdensity of 200, so need to be multiplied by about 1.19 to compare the other virial masses cited, see \citet{Bland_16}. With this correction, their range covers our value.

Stellar streams are powerful probes of the MW gravitational potential \citep{Johnston_99}.
Mainly from the precession of the Sagittarius stream, \citet{Gibbons_14} derive low masses for the MW at 50 and 100 kpc, $M_\mathrm{50}=0.29\pm0.05\times10^{12}$M$_\odot$ and $M_\mathrm{100}=0.41\pm0.07\times10^{12}$M$_\odot$.  This method is not free of assumptions,  especially regarding the form of the underlying potential:  \citet{Fardal_19} use similar properties of the Sagr stream but different potential slopes and shapes for the halo of the MW to obtain a preferred mass $M_\mathrm{100}=0.70\times10^{12}$M$_\odot$. \citet{Erkal_18} use the Orphan stream to obtain a mass of $M_\mathrm{50}=0.39\pm0.02\times10^{12}$M$_\odot$. 
Their model includes the influence of the LMC, although the shape of the LMC potential due to the infall onto the MW is not taken into account; furthermore, the dark matter wake introduced by the gravitational influence of the LMC might cause deviations of the potential of the MW from a prolate/oblate NFW profile \citep{Garavito_19}. 

Globular clusters are rather abundant in the inner halo and thus a useful tracer of those regions. \citet{Watkins_18} used halo globular clusters with proper motions from Gaia DR2 \citep{Brown_18,Helmi_18} and HST \citep{Sohn_18}, to obtain a value of $M_\mathrm{39.5}=0.42^{+0.07}_{-0.06}\times10^{12}$M$_\odot$ using their Wa10 mass estimator. 
 \citet{Vasiliev_19} use \textit{Gaia} DR2 to measure the proper motion of globular clusters out to $\sim$100 kpc, and  they then use distribution functions to obtain $M_\mathrm{50}=0.54^{+0.11}_{-0.07}\times10^{12}$M$_\odot$ and $M_\mathrm{100}=0.85^{+0.33}_{-0.20}\times10^{12}$M$_\odot$. 
\citet{Eadie_19} use the data sets of \citet{Sohn_18}, \citet{Helmi_18}, \citet{Vasiliev_19} outside of 15 kpc and the method of \citet{Eadie_15}
to obtain $M_\mathrm{50}=0.37^{+0.06}_{-0.04}
\times10^{12}$M$_\odot$ and $M_\mathrm{100}=0.53^{+0.09}_{-0.07}\times10^{12}$M$_\odot$. The difference between the last works show that even when the same proper motions are used, other assumptions,  such as  the radial distribution of the tracer population, can change the obtained value of the MW mass.

In the past the mass obtained using dwarf galaxies often depended on whether or not Leo~I was assumed to be a bound satellite, see e.g. \citet{Kulessa_92}. When assumed to be bound, its HST proper motion leads to a mass of $M_\mathrm{vir}=1.6\pm0.4\times10^{12}$M$_\odot$
 \citep{Boylan-Kolchin_13}. That is slightly higher than our estimate but consistent at the 1$\,\sigma$ level. 

\citet{Watkins_10} obtain a preferred estimate of  $M_\mathrm{300}=2.7\times10^{12}$M$_\odot$, using the Wa10 mass estimator on a sample of 6 dwarf galaxies with and 20 without proper motions  as was available in the data at that time. They found a $\beta=-4.5$.
The recently improved determination of $\beta$, which is still tangential but less  than the value reported in \citet{Watkins_10}, leads to a lower mass estimate. Our analysis also suggests that the
Wa10 value was also relatively high due to the large value of $\gamma$ coming  from an incomplete sample of dwarfs,  and by the assumption that the mass estimator is unbiased when applied to samples in which galaxies close to their pericenters are overrepresented.
  
Using HST proper motions of classical dwarfs and the  LMC along with the Illustris simulation to match specific angular momenta of the observed dwarfs and their analogues in the simulation,
\citet{Patel_18} obtain $M_\mathrm{vir}=0.96^{+0.29}_{-0.28}\times10^{12}$M$_\odot$.
Matching classical dwarf galaxy distribution functions, constructed using HST and \textit{Gaia} proper motions, and the EAGLE simulations using a distribution function by fitting  angular momenta and specific energy, \citet{Callingham_19} obtained $M_\mathrm{vir,200}=1.17^{+0.21}_{-0.15}\times10^{12}$M$_\odot$. 
Upscaled to $M_\mathrm{vir}$ results in $\approx1.39\times10^{12}$M$_\odot$.
Very recently \citet{Li_19b} used a similar sample of dwarf galaxies and a distribution function method based on the the EAGLE simulation to obtain $M_\mathrm{vir,200}=1.23^{+0.21}_{-0.18}\times10^{12}$M$_\odot$. Upscaled to $M_\mathrm{vir}$ this converts in $\approx1.46\times10^{12}$M$_\odot$,  well matching our result.

In summary, there is no clear trend of low or high masses depending on the particular tracer, with each being used to recover a range of masses, although it can be noted that the masses obtained from streams do not cover the high mass tail of the various estimates. Our estimated mass  of $M_\mathrm{vir}=1.51^{+0.45}_{-0.40} \times 10^{12}M_{\odot}$ is intermediate between literature estimates (where the largest one is from the work by \citealt{Boylan-Kolchin_13} and the smallest one by \citealt{Kafle_14}). 

The  way that the LMC is accounted for (or not accounted for) could explain at least part of the differences in the various mass estimates. It is likely that our MW mass estimate includes the mass of the LMC at least partly, since the majority of our tracers are at larger distances than the LMC and thus for them the force of the LMC acts in about the same direction as the force of the MW. The LMC alone is more massive (about 20\% of the MW mass) than a median subhalo population in ELVIS, which amount to just 10\% of the main halo. Thus, the LMC is more important than the typical subhalos.
The LMC might also explain why our mass at 64 kpc, $0.58^{+0.15}_{-0.14}\times10^{12}$M$_\odot$, is larger than the mass by \citet{Erkal_18} who included the LMC explicitly. If our virial mass includes the LMC our results would then fit better to the timing result of \citet{Penarrubia_16}, who obtain $M_\mathrm{MW}=1.04^{+0.26}_{-0.23}\times10^{12}$M$_\odot$ and $M_\mathrm{LMC}=0.25^{+0.09}_{-0.08}\times10^{12}$M$_\odot$.

%%%%%%%%%%%%%%%%%%%%%%%%%%%%%%
\section{Conclusions and summary} \label{sec:conclusions}
%%%%%%%%%%%%%%%%%%%%%%%%%%%%%%%%%%%%

The mass of the Milky Way is still surprisingly uncertain, notwithstanding the impressive amount of literature on the topic and the variety of approaches taken. Due to their large distances, dwarf galaxies are a promising tracer to sample the mass of the MW to its outskirts. Thanks to \textit{Gaia} DR2, there are now systemic proper motions and line-of-sight velocities available for 45 galaxies within 280 kpc \citep{Fritz_18a,Fritz_18b,Helmi_18,Carlin_18,Torrealba_18b, Longeard_19}; this leverages the mass determination by also providing information on the velocity anisotropy of the tracer population. 

We use this \textit{Gaia} DR2 data set to estimate the mass of the Milky Way using the scale free mass estimator of \citet{Watkins_10}. We determine the potential
index $\alpha$, the anisotropy parameter $\beta$, and the tracer density index $\gamma$ parameter for the Milky Way.
We determine, from a likely complete sample (15 galaxies with M$_V<-5.9$), that $\gamma=2.11\pm0.23$, but we note that the galaxy number density profile is not well fit by a single power-law, with a deficit of satellites within 18 kpc. 
We obtain $\beta$=$-0.67^{+0.45}_{-0.62}$, or $\beta$=$-0.21^{+0.37}_{-0.51}$ when we exclude from the sample all possible former satellites of the LMC. We determine $\alpha$ iteratively, combining a direct calculation of observed mass within 19 kpc from circular velocity determinations and estimates of the mass within 273\,kpc from the Wa10 estimator. This implies that the value of $\alpha$ depends on the MW mass. For our derived Milky Way mass ($M_\mathrm{273}$) and its uncertainty, it follows that $\alpha=0.27^{+0.12}_{-0.11}$.

We use cosmological simulations to check for biases introduced by  observational uncertainties and by the radial distribution of tracers.
We find that the mass obtained is biased even without observational uncertainties. This appears to be the result of the distribution of subhalos not being scale-free, as assumed in the Wa10 estimator, but having a deficit of satellites in the inner region where galaxies are more likely close to their pericenters. 
We correct for that bias by applying a correction factor, which is included in the following listed results.
\\

Summarising our main results:

\begin{itemize}
\item  Observational uncertainties lead to an increase in the value of the recovered mass with, as expected, larger errors. We devise a method to identify which galaxies introduce the largest uncertainties, due to their proper motion errors, and exclude them, resulting in a sample of 36 galaxies. 
\item We use different samples to assess the importance of former satellites of the LMC and reference frames to account for the reflex motion of the MW due to the LMC.
\item Using the full radial range of galaxies, we obtain a mass of $1.43^{+0.35}_{-0.32}\times10^{12}$M$_\odot$ within 273 kpc. This is consistent with the value determined when using only the outer half of the sample. 
\item From the inner half of the sample we obtain $0.58^{+0.15}_{-0.14}\times10^{12}$M$_\odot$ within 64 kpc.
\item We derive a virial mass of $M_\mathrm{vir}=1.51^{+0.45}_{-0.40} \times 10^{12}M_{\odot}$ within R$_\mathrm{vir}=308\pm29$ kpc for an overdensity of 97, by averaging the various cases (i.e. with or without former LMC satellites). 
\item We find that, of the scaling parameters, the mass profile $\alpha$ has the biggest impact on the mass. When combined, the three scaling parameters have nearly the same importance as the mass estimator calibration and the intrinsic uncertainty.
\end{itemize}

The mass determination from this work is slightly larger than the average of literature values given by \citet{Bland_16}, 
and thus provides support for an intermediate mass for the Milky Way \citep{Fritz_18a,Watkins_18}, 
disfavouring a low mass Milky Way solution to the `too big to fail' (TBTF) problem \citep{Boylan-Kolchin_11b,DiCintio11,Wang_12a,Vera-Ciro_13},  indicating the need for other solutions such as feedback \citep[e.g.][]{Brook15,Wetzel_16} or modifications to CDM \citep[e.g.][]{Lam_17}. 

Adopting a Milky Way stellar mass of 6$\times 10^{10}$M$_{\odot}$ from  \cite{Bland_16}, we can compare to abundance matching results: our inferred virial mass places the Milky Way at the expected position in the $\rm M^*$-$\rm M_{\rm halo}$ relation of \citet{Kravtsov18}, while, interestingly, all other abundance matching studies would predict a larger value for the Milky Way mass (see \citealt{Behroozi19,Somerville18,Wechsler2018,RP17} for review).

Our result can also be compared with the mass of M31, for which derived masses range   from about  $M_\mathrm{vir}=0.8\pm0.1\times10^{12}$M$_\odot$ \citep{Kafle_18}, over $M_\mathrm{M31\,300}= 1.40\pm0.43\times10^{12}$M$_\odot$ \citep{Watkins_10} to  $M_\mathrm{vir}=2.1\pm0.5\times10^{12}$M$_\odot$ \citep{Fardal_13}. Our MW mass is slightly less than half the total mass of the Local Group, $M_\mathrm{LG}=3.17\pm0.57\times10^{12}$M$_\odot$ as derived by \citet{Marel_12}, but slightly larger than half the estimate by \citet{Penarrubia_16} of $M_\mathrm{LG}=2.64^{+0.42}_{-0.38}\times10^{12}$M$_\odot$. Thus, the question of which galaxy is the most massive, between the MW and M31, remains open. For both galaxies there is the complication that there is a medium massive galaxy closeby (LMC for MW, M33 for M31) which is likely partly included in $M_\mathrm{vir}$ estimates. On the one hand M33 is probably more massive, since it is more luminous \citep{McConnachie_12}, on the other hand the LMC is closer to its host galaxy, and thus probably has a greater  influence on mass estimates. 
Improved measurements may answer the question of which galaxy is the most massive  in the Local Group. 

While \textit{Gaia} will improve its precision, the sample of satellites will maintain a bias towards those close to the Milky Way due to the rather bright detection limit. This is true for the Milky Way and even more so for M31: at its distance, all old stars are fainter than the detection limit of \textit{Gaia}. 

To obtain  proper motions of more distant satellites other measurements are needed, in order to determine their 3-dimensional dynamics: these are in progress with HST \citep{Kallivayalil_15,Kallivayalil_16,Weisz_19}.

%%%%%%%%%%%%%%%%%%%%%%%%%%%%%%%%%%%%%%%%%%%%%%%%%%%
\section*{Acknowledgements}
%%%%%%%%%%%%%%%%%%%%%%%%%%%%%%%%%%%%%%%%%%%%%%%%%%%
 The authors thank the referee for the thorough and constructive report, which significantly helped to improve the manuscript. TKF, GB, CB, and ST acknowledge financial support through the grants (AEI/FEDER, UE) AYA2017-89076-P, AYA2016-77237-C3-1-P and AYA2015-63810-P, as well as by the Ministerio de Ciencia, Innovaci\'{o}n y Universidades (MCIU), through the State Budget and by the Consejer a de Economia, Industria, Comercio y Conocimiento of the Canary Islands Autonomous Community, through the Regional Budget.
ADC acknowledges financial support from a Marie-Sk\l{}odowska-Curie Individual Fellowship grant, H2020-MSCA-IF-2016 Grant agreement 748213 DIGESTIVO. 
 GB gratefully acknowledges financial support from Spanish Ministry of
Economy and Competitiveness (MINECO) under the Ramon y Cajal Programme
(RYC-2012-11537). 
 CB is supported by a MCIU Ram\'{o}n y Cajal Fellowship (RYC 2013-12784).
We thank Laura Watkins who provided data to test the mass estimator in the idealized case. We thank Mike Boylan-Kolchin for providing the 1$\,\sigma$ mass interval of \citet{Boylan-Kolchin_13} and Gwendolyn Eadie  for providing the 1$\,\sigma$ mass intervals of \citet{Eadie_19}.

%%%%%%%%%%%%%%%%%%%%%%%%%%%%%%%%%%%%%%%%%%%%%%%%%%

\bibliographystyle{mnras}
\bibliography{archive} 

%%%Apendix
\appendix
\section{Details on sample}
\label{sec:ap}

We list in Table~\ref{KapSou3} the sources of the used properties like distance modulus, proper motion and line-of-sight velocity. For further details, see \citet{Fritz_18a}.

      \begin{table*}
      \caption[]{Sample of galaxies. Col. 1 lists the object name;  in col. 2 we provide the distance moduli (d.m.) used and their source; in col. 3 we provide the line of sight velocities and their source ;  in col. 4/5 we list  the proper motions and their sources in both dimension and in col. 6 the systematic uncertainties of both.}
         \label{KapSou3}
     $$
         \begin{array}{p{0.13\linewidth}lllll}
            \hline
            name   & \mathrm{d.m.} & \mathrm{v}_{\mathrm{l.o.s.}}  & \mu_\alpha &  \mu_\delta & \mathrm{syst.~} \mu \mathrm {~uncertainty} \\
            \hline
     &  & \mathrm{km/s} & \mathrm{[mas/yr]} & \mathrm{[mas/yr]} & \mathrm{[mas/yr]} \\         
            \hline
 Antila II &  20.60\pm0.11^{ 77} &291^{ 77} &-0.095\pm0.187^{ 77} &0.058\pm0.024^{ 77} & 0.035^{ \mathrm{This ~paper}} \\
 Aquarius II &  20.16\pm0.07^{ 1} &-71^{ 1} &-0.252\pm0.526^{ 72} &0.011\pm0.448^{ 72} & 0.063^{ 72} \\
 Bo\"{o}tes I & 19.11\pm0.08^{ 32} &99^{ 2, 30} &-0.554\pm0.092^{ 72} &-1.111\pm0.068^{ 72} & 0.035^{ 72} \\
 Bo\"{o}tes II &  18.11\pm0.06^{ 33} &-117^{ 3} &-2.686\pm0.389^{ 72} &-0.53\pm0.287^{ 72} & 0.056^{ 72} \\
 Bo\"{o}tes III &  18.35\pm0.1^{ 74} &198^{ 73} &-1.14\pm0.18^{  75} &-0.98\pm0.2^{  75} & 0.035^{ \mathrm{This~paper}} \\
 Canes Venatici I & 21.62 \pm 0.05^{ 34} &31^{ 2, 4} &-0.159\pm0.094^{ 72} &-0.067\pm0.054^{ 72} & 0.035^{ 72} \\
 Canes Venatici II &  21.02 \pm 0.06^{ 35} &-129^{  4} &-0.342\pm0.232^{ 72} &-0.473\pm0.169^{ 72} & 0.056^{ 72} \\
 Carina I &  20.08\pm  0.08^{ 57, 58} &229^{ 5} &0.485\pm0.017^{ 72} &0.131\pm0.016^{ 72} & 0.035^{ 72} \\
 Carina II &  17.79\pm0.05^{ 36} &477^{ 6} &1.867\pm0.078^{ 72} &0.082\pm0.072^{ 72} & 0.035^{ 72} \\
 Carina III & 17.22\pm0.10^{ 36} &285^{ 6} &3.046\pm0.119^{ 72} &1.565\pm0.135^{ 72} & 0.057^{ 72} \\
 Columba II &  21.31\pm0.11^{ 53} &156^{ 76} &0.33\pm0.28^{ 76} &-0.38\pm0.38^{ 76} & 0.062^{ 76} \\
 Coma Berenices I &  18.13\pm 0.08^{ 37} &98^{ 4} &0.471\pm0.108^{ 72} &-1.716\pm0.104^{ 72} & 0.035^{ 72} \\
 Crater II &  20.25\pm0.10^{ 39} &88^{ 9} &-0.184\pm0.061^{ 72} &-0.106\pm0.031^{ 72} & 0.035^{ 72} \\
  Draco I &  19.49\pm 0.17^{ 59, 60} &-291^{ 10} &-0.012\pm0.013^{ 72} &-0.158\pm0.015^{ 72} & 0.035^{ 72} \\
 Draco II &  16.66\pm 0.04^{ 40} &-348^{ 11} &1.242\pm0.276^{ 72} &0.845\pm0.285^{ 72} & 0.057^{ 72} \\
 Fornax I &  20.72\pm0.04^{ 61} &55^{ 5, 13} &0.374\pm0.004^{ 72} &-0.401\pm0.005^{ 72} & 0.035^{ 72} \\
  Grus I &  20.4\pm0.2^{ 44} &-141^{ 14} &-0.261\pm0.172^{ 72} &-0.437\pm0.238^{ 72} & 0.046^{ 72} \\
 Hercules I &  20.64\pm0.14^{ 42, 43} &45^{ 4, 15} &-0.297\pm0.118^{ 72} &-0.329\pm0.094^{ 72} & 0.035^{ 72} \\
 Horologium I & 19.46\pm0.2^{ 44, 45} &169^{ 16} &1.52\pm0.25^{ 72} &-0.47\pm0.39^{ 72} & 0.058^{ 72} \\
  Horologium II & 19.6\pm0.2^{ 78} &113^{ 76} &0.891\pm0.088^{ 76} &-0.55\pm0.08^{ 76} & 0.049^{ 76} \\
 Hydra II &  20.89\pm 0.12^{ 46} &303^{ 7} &-0.416\pm0.519^{ 72} &0.134\pm0.422^{ 72} & 0.061^{ 72} \\
 Hydrus I &  17.20 \pm0.04^{ 29} &80^{ 29} &4.044\pm0.312^{ 72} &-1.755\pm0.276^{ 72} & 0.035^{ 72} \\
 Leo I &  22.15\pm 0.1^{ 62} &283^{ 17} &-0.086\pm0.059^{ 72} &-0.128\pm0.062^{ 72} & 0.035^{ 72} \\
 Leo II &  21.76\pm 0.13^{ 63, 64} &78^{ 18, 31} &-0.025\pm0.08^{ 72} &-0.173\pm0.083^{ 72} & 0.035^{ 72} \\
 Leo IV &  20.94 \pm 0.07^{ 47} &132^{ 4} &-0.59\pm0.531^{ 72} &-0.449\pm0.358^{ 72} & 0.059^{ 72} \\
 Leo V & 21.19 \pm 0.06^{ 48} &173^{ 19} &-0.097\pm0.557^{ 72} &-0.628\pm0.302^{ 72} & 0.057^{ 72} \\
 LMC &  18.50\pm0.02^{ 80} &262^{ 83} &1.85^{ 82} &0.234^{ 82} & 0.030^{ 82} \\
 SMC &  18.99\pm0.12^{ 80} &146^{ 80,84} &0.797^{ 82} &-1.22^{ 82} & 0.030^{ 82} \\
 Pisces II &  21.31\pm0.18^{ 49} &-227^{ 7} &-0.108\pm0.645^{ 72} &-0.586\pm0.498^{ 72} & 0.061^{ 72} \\
 Phoenix II &  19.60\pm0.1^{ 50} &33^{ 76} &0.5\pm0.12^{ 76} &-1.16\pm0.14^{ 76} & 0.059^{ 76} \\
 Reticulum II &  17.5\pm0.1^{ 50} &63^{ 21} &2.398\pm0.04^{ 72} &-1.319\pm0.048^{ 72} & 0.035^{ 72} \\
 Reticulum III &  19.82\pm0.31^{ 79} &274^{ 76} &-0.39\pm0.53^{ 76} &-0.32\pm0.63^{ 76} & 0.058^{ 76} \\
 Sagittarius I &  17.13\pm0.11^{ 65} &140^{ 80} &-2.736\pm0.009^{ 72} &-1.357\pm0.008^{ 72} & 0.035^{ 72} \\
 Sagittarius II &  19.32\pm0.03^{ 81} &-177^{ 81} &-0.65\pm0.09^{ 81} &-0.88\pm0.12^{ 81} & 0.035^{ 85} \\
 Sculptor I &  19.64\pm 0.13^{  67, 68} &111^{ 5, 13} &0.084\pm0.006^{ 72} &-0.133\pm0.006^{ 72} & 0.035^{ 72} \\
 Segue 1 &  16.8\pm 0.2^{ 51} &209^{ 22} &-1.697\pm0.195^{ 72} &-3.501\pm0.175^{ 72} & 0.035^{ 72} \\
 Segue 2 & 17.8\pm0.18^{ 52} &-39^{ 23} &1.656\pm0.155^{ 72} &0.135\pm0.104^{ 72} & 0.045^{ 72} \\
 Sextans I &  19.67\pm 0.15^{ 68} &224^{ 24} &-0.438\pm0.028^{ 72} &0.055\pm0.028^{ 72} & 0.035^{ 72} \\
 Triangulum II &  17.27\pm0.1^{ 53} &-382^{ 25} &0.588\pm0.187^{ 72} &0.554\pm0.161^{ 72} & 0.051^{ 72} \\
 Tucana II &  18.8\pm 0.2^{ 44, 45} &-129^{ 14} &0.91\pm0.059^{ 72} &-1.159\pm0.074^{ 72} & 0.035^{ 72} \\
 Tucana III &  16.8\pm 0.1^{ 50} &-102^{ 26, 27} &-0.025\pm0.034^{ 72} &-1.661\pm0.035^{ 72} & 0.035^{ 72} \\
 Ursa Major I &  19.94\pm 0.13^{ 54} &-55^{ 2, 4} &-0.683\pm0.094^{ 72} &-0.72\pm0.13^{ 72} &  0.035^{ 72} \\
 Ursa Major II &  17.70\pm 0.13^{ 55} &-117^{ 2, 4} &1.691\pm0.053^{ 72} &-1.902\pm0.066^{ 72} & 0.035^{ 72} \\
 Ursa Minor I &  19.40\pm 0.11^{ 70, 71} &-247^{ 28} &-0.184\pm0.026^{ 72} &0.082\pm0.023^{ 72} & 0.035^{ 72} \\
 Willman 1 &  17.90\pm0.40^{ 56} &-12^{ 2} &0.199\pm0.187^{ 72} &-1.342\pm0.366^{ 72} & 0.051^{ 72} \\     
            \noalign{\smallskip}
            \hline
         \end{array}
$$
  \tabletext{(1) \citet{Torrealba_16a};
(2) \citet{Martin_07}; (3) \citet{Koch_09}; (4) \citet{Simon_07};
(5) \citet{Walker_09_clas}; (6) \citet{Li_18a}; (7) \citet{Kirby_15};
 (9) \citet{Caldwell_17}; (10) \citet{Walker_15};
(11) \citet{Martin_16};  (13) \citet[][and references therein]{BattagliaSt_12}; (14) \citet{Walker_16}; (15) \citet{Aden_09}; (16) \citet{Koposov_15b}; (17) \citet{Mateo_08}; (18)  \citet{Spencer_17}; (19) \citet{Walker_09}; (21) \citet{Simon_15}; (22) \citet{Simon_11}; (23) \citet{Kirby_13}; (24) \citet{Cicuendez_18}; (25) \citet{Kirby_17}; (26) \citet{Simon_17}; (27) \citet{Li_18b}; (28) \citet{Walker09}, (29) \citet{Koposov_18}; (30) \citet{Koposov_11}; (31) \citet{Koch_07}; (32) \citet{DallOra_06}; (33) \citet{Walsh_08}; (34) \citet{Kuehn_08}; (35) \citet{Greco_08}; (36) \citet{Torrealba_18}; (37) \citet{Musella_09}; (38) \citet{Weisz_16}; (39) \citet{Joo_18}; (40) \citet{Longeard_18}; (42) \citet{Musella_12}; (43) \citet{Garling_18}; (44) \citet{Koposov_15a}; (45) \citet{Bechtol_15}; (46)\citet{Vivas_16}; (47) \citet{Moretti_09}; (48) \citet{Medina_17}; (49) \citet{Sand_12}; (50) \citet{Mutlu_18}; (51) \citet{Belokurov_07}; (52) \citet{Boettcher_13}; (53) \citet{Carlin_17}; (54) \citet{Garofalo_13}; (55) \citet{DallOra_12}; (56)  \citet{Willman_06}, (57) \citet{Coppola_15}; (58) \citet{Vivas_13}; (59) \citet{Bonanos_04}; (60) \citet{Kinemuchi_08}; (61) \citet{Rizzi_07}; (62) \citet{Stetson_14}; (63) \citet{Bellazzini_05}; (64) \citet{Gullieuszik_08}; (65) \citet{Hamanowicz_16}; (67) \citet{Martinez_16}; (68) \citet{Pietrzynski_08}; (69) \citet{Mateo_95}; (70) \citet{Carrera_02}; (71) \citet{Bellazzini_02}; (72) \citet{Fritz_18a}; (73) \citet{Carlin_09}; (74) \citet{Grillmair_09}; (75) \citet{Carlin_18}; (76) \citet{Fritz_18b}; (77) \citet{Torrealba_18b}; (78) \citet{Kim_15a}; (79) \citet{Drlica-Wagner_15}; (80) \citet{McConnachie_12}; (81) \citet{Longeard_19}; (82) \citet{Helmi_18}; (83) \citet{Alves_04}; (84) \citet{Graczyk_14}.
}
   \end{table*}

%%%%%%%%%%%%%%%%%%%%%%%%%%%%%%%%%%%%%%%%%%%%%%%%%%

% Don't change these lines
\bsp	% typesetting comment
\label{lastpage}
\end{document}